\documentclass[review]{elsarticle}
\DeclareUnicodeCharacter{2219}{\textbullet}
\makeatletter
\def\ps@pprintTitle{%
 \let\@oddhead\@empty
 \let\@evenhead\@empty
 \def\@oddfoot{\centerline{\thepage}}%
 \let\@evenfoot\@oddfoot}
\makeatother

\biboptions{numbers,sort&compress}
\pdfoutput=1
\usepackage{graphicx}
\usepackage{amsfonts, amsmath, amssymb}
\usepackage{mathabx}
\usepackage{booktabs,siunitx}
\usepackage[svgnames,table]{xcolor}
\usepackage[tableposition=above]{caption}
\usepackage{pifont}
\usepackage{bm}
\usepackage{cancel}
\usepackage{caption}
\usepackage{color}
\usepackage{enumerate}
\usepackage{float}
\usepackage{hyperref}
\usepackage{soul}
\usepackage[english]{babel}
\usepackage[margin=1in]{geometry}
\usepackage{mathtools}
\usepackage{multirow}
\usepackage{setspace}
\usepackage{subfigure}
\usepackage{stmaryrd}
\usepackage{lineno}
\DeclareUnicodeCharacter{00B3}{\textsuperscript{3}}
\usepackage[ruled,linesnumbered]{algorithm2e}
\RestyleAlgo{ruled}
\SetKwComment{Comment}{/* }{ */}

\SetCommentSty{mycommfont}
\DeclareMathAlphabet{\mathpzc}{OT1}{pzc}{m}{it}

\renewcommand \d[1]{{\rm{d}} #1}
\newcommand \D [2]{\frac{\partial #1}{\partial #2}}

\renewcommand{\vec}[1]{\bm{\mathrm{#1}}}
\newcommand{\V}[1]{\bm{\mathrm{#1}}}

\newcommand{\hglt}[1]{\textcolor{black}{#1}}

\def \div{\nabla \cdot \mbox{}}

\def \bnabla{\V{\nabla}}

\def \x{\vec{x}}
\def \e{\vec{e}}

\def \r{\vec{r}}

\def \v{\vec{v}}
\def \vone{\vec{v}_1}
\def \pone{p_1}
\def \vpone{\vec{p}_1}
\def \vptwo{\vec{p}_2}

\def \rhoone{\rho_1}
\def \vonehat{\widehat{\vec{v}}_1}
\def \ponehat{\widehat{p}_1}
\def \rhoonehat{\widehat{\rho}_1}
\def \ahat{\widehat{a}}
\def \bhat{\widehat{b}}
\def \v2{\vec{v}_2}

\def \e{\vec{e}}
\def \iota{\imath}

\def \A{\vec{A}}
\def \b{\vec{b}}
\def \bv{\vec{b_v}}

\def \rv{\vec{r_v}}
\def \rp{\vec{r_p}}

\def \G{\vec{G}}

\def \Lmu{\vec{L_{\mu}}}
\def \Drho{\vec{D_{\rho}}}
\def \Llamb{\vec{L_{\lambda}}}
\def \vrho{\vec{\rho}}

\def \v{\vec{v}}
\def \vvarphi{\vec{\varphi}}

\def \x{\vec{x}}

\def \ev{\vec{e_v}}
\def \ep{\vec{e_p}}

\def \div{\nabla \cdot \mbox{}}

\def \dx{\Delta x}
\def \dy{\Delta y}

\def \presdisp{\vec{d}_1}
\def \pfunc{p}
\def \half{\frac{1}{2}}
\def \3half{\frac{3}{2}}

\def \etal{\emph{et al.}}

\newcommand{\upperRomannumeral}[1]{\uppercase\expandafter{\romannumeral#1}}

\begin{document}
\let\today\relax
\let\underbrace\LaTeXunderbrace
\let\overbrace\LaTeXoverbrace

\begin{frontmatter}
	
\title{Consistent continuum equations and numerical benchmarks for a perturbation-based, variable-coefficient acoustofluidic solver}

\author[UNL]{Khemraj Gautam Kshetri\fnref{eq_contrib}}
\author[SDSU]{Amneet Pal Singh Bhalla\corref{mycorrespondingauthor}\fnref{eq_contrib}}
\ead{asbhalla@sdsu.edu}
\author[UNL]{Nitesh Nama\corref{mycorrespondingauthor}}
\ead{nitesh.nama@unl.edu}

\address[UNL]{Department of Mechanical Engineering, University of Nebraska Lincoln, Lincoln, NE}
\address[SDSU]{Department of Mechanical Engineering, San Diego State University, San Diego, CA}
\cortext[mycorrespondingauthor]{Corresponding author}
\fntext[eq_contrib]{These authors contributed equally}

\begin{abstract}
We present a consistent continuum framework and a variable-coefficient acoustofluidic solver to analyze acoustic streaming. A perturbation approach is used to split the compressible Navier-Stokes equations into two sub-systems: a first-order harmonic system and a time-averaged second-order mean system. Prior acoustofluidic numerical studies have typically employed simplifying assumptions regarding the second-order mass balance equation and boundary conditions. These assumptions---frequently left unjustified---result in an incongruent problem statement where the boundary conditions are not consistent with the governing equations. To clarify these assumptions and mitigate the associated confusion, we systematically formulate the second-order mass balance equation into two analytically equivalent but numerically distinct forms by introducing the fluid's Lagrangian and mass transport velocity. We clarify the relation between these quantities to illustrate that a zero mass transport velocity does not necessarily imply a zero Lagrangian velocity. A second-order accurate finite difference/volume scheme is used to discretize the governing equations and boundary conditions expressed in Eulerian form. We solve the first-order coupled Helmholtz system via a sparse direct solver while the second-order system is solved via a Krylov (FGMRES) solver with a projection method-based preconditioner. To ascertain the spatial accuracy and consistency of numerical implementation of the split system of equations and boundary conditions, we demonstrate verification results and convergence rates using manufactured solutions. We provide benchmark test cases to demonstrate that the choice of boundary conditions and the oscillation profile of the microchannel wall can, in general, strongly impact the second-order flow field. Lastly, we examine a system with spatially varying density to illustrate the difference in flow fields between constant and variable density systems.

\end{abstract}


\end{frontmatter}

\section{Introduction}

In recent years, acoustofluidics has emerged as an innovative technology for a variety of lab-on-a-chip applications~\cite{laurell2014microscale,ding2013surface,durrer2022robot,wei2023microscale}. In acoustofluidic systems, high-frequency acoustic waves interact nonlinearly with viscous fluids. The result of these interactions is not just a high-frequency oscillatory response of the fluid, but also a mean flow called acoustic streaming. Acoustic streaming has recently been used to enable fast, microscale fluid mixing and pumping---two tasks that are otherwise difficult to accomplish at small scales because of the low Reynolds number~\cite{huang2015acoustofluidic,ahmed2013tunable,junahuang2014reliable,huang2013acoustofluidic,ozcelik2014acoustofluidic,pavlic2023sharp}. 

While acoustic streaming has been a subject of scientific investigations dating back to the $18^\textrm{th}$ century, its merger with microfluidics, and associated applications for lab-on-a-chip systems, has led to a renewed interest in understanding nonlinear fluid-acoustic interactions~\cite{beyer1999sounds,friend2011microscale}. There have been several numerical studies recently to model microscale acoustofluidic phenomena~\cite{nama2014investigation,nama2015numerical,muller2013ultrasound,devendran2022role,vanneste2011streaming,orosco2022modeling,lei2017transducer,baasch2018acoustofluidic,sankaranarayanan2008flow}. Nyborg's perturbation approach is typically employed in these studies to separate the fluid response into a first-order oscillating and a second-order mean component. As a result of this approach, there are two sets of linear equations: the first-order system describing the fluid's oscillatory response, and the second-order system describing its mean response. 

Considering that fluid compressibility is a prerequisite for acoustic wave propagation, most of these studies begin with compressible Navier-Stokes equations. The perturbation expansion of compressible Navier-Stokes equations results in a time-averaged mass source term for the second-order system. Since most acoustofluidic systems are characterized by low Mach numbers (i.e., with characteristic streaming velocities that are small comparable to the speed of sound), the second-order mass source is sometimes overlooked~\cite{lei2017transducer,sankaranarayanan2008flow,devendran2014separation}. While this is not problematic in isolation, neglecting a mass source while employing a non-zero Eulerian velocity boundary condition at the oscillating boundaries of the computational domain can, in general, result in situations where the boundary conditions are inconsistent with the governing equations ~\cite{baasch2020acoustic}. However, for certain types of wall actuation, the mass source term has little impact and numerical solvers often produce seemingly-correct solutions despite this inconsistency. Alternatively, some studies retain the mass source term, but prescribe zero-velocity boundary condition at the mean position of the oscillating boundary~\cite{nama2015numerical,muller2012numerical,das2019acoustothermal,ghorbani2022acoustic}. Again, this can, in general, lead to an incongruent problem statement if the boundary conditions are not in agreement with the governing equations.

A related issue in modeling acoustofluidic systems is determining appropriate boundary conditions for the oscillating boundary of the second-order system. In previous studies, different boundary conditions were prescribed for the mean flow at the oscillating wall. In some studies, the oscillating wall is assigned a zero Eulerian velocity~\cite{nama2015numerical,muller2012numerical,das2019acoustothermal,ghorbani2022acoustic}, while in others, a zero mass transport velocity~\cite{das2022extra,muller2014numerical,muller2015theoretical} or a zero Lagrangian velocity is prescribed~\cite{kshetri2023acoustophoresis,kshetri2024evaluating,baasch2019acoustic,pavlic2022influence}. In general, these velocities represent distinct quantities that are not identical. While this issue was discussed by Bradley~\etal~\cite{bradley1996acoustic}, some subsequent studies that neglected this distinction still produced results that matched experimental observations~\cite{muller2013ultrasound,nama2016investigation}; this has resulted in proliferation of numerical models where contradictory boundary conditions continue to be employed and justified by referencing prior numerical reports. A lack of rigorous benchmarking problems in computational acoustofluidic studies further exacerbates this confusion. In most studies, the numerical implementations are benchmarked against  Muller~\etal~\cite{muller2012numerical,muller2013ultrasound}. Acoustofluidic setups discussed in~\cite{muller2012numerical,muller2013ultrasound} consider uniform rectilinear oscillations of the channel boundary, and are insensitive to the inconsistencies discussed above. As such, the applicability of a numerical implementation, compared against Muller~\etal~\cite{muller2012numerical,muller2013ultrasound}, remains unclear for general oscillations of the boundary. Further,
the validity of numerical acoustofluidic models is often assessed by how well it reproduces previous numerical results from the literature. The accuracy of the implementation and convergence rates associated with it are rarely reported using formal numerical implementation verification techniques such as the method of manufactured solutions. As noted above, for certain wall oscillation profiles and parameter ranges, the numerical solver can converge despite the inherent inconsistencies in the problem statement. As such, these inconsistencies---that can be highlighted through formal numerical verification techniques---remain undiscussed leading to unclear implications concerning the general applicability of these models. We aim to provide a quantitative benchmark for test cases that are sensitive to boundary conditions in this work. 


Most numerical studies on acoustofluidic systems assume a spatially constant density/viscosity of the actuated fluid. However, recent experimental investigations have reported boundary-driven streaming in inhomogeneous fluids and the associated relocation of density/compressibility inhomogeneities into stabilized configurations~\cite{deshmukh2014acoustic}. These observations were later theoretically and numerically explained by Karlsen~\etal~\cite{karlsen2016acoustic,karlsen2018acoustic}. Nonetheless, the specific device configuration used in these reports was again characterized by uniform rectilinear oscillations of the boundary. As such, no numerical benchmark exists for a variable coefficient solver for non-uniform boundary actuations. In this paper, we present a formally-verified acoustofluidic solver and leverage it to provide benchmark problems for variable coefficient acoustofluidic solvers that deal with non-uniform, non-rectilinear oscillations of the channel boundary. This is an essential first step towards validating a multiphase/multicomponent acoustofluidic solver that can enable numerical simulations of
the transport of gas bubbles (with surface tension forces at the gas-liquid interface) and biological cells (with elastic forces at the liquid-structure interface) within fluid-filled microfluidic devices. 

We revisit the perturbation approach with a specific focus on deriving consistent continuum equations and boundary conditions for a variable-coefficient acoustofluidic solver. Beginning with the compressible Navier-Stokes equations, we systematically derive the first- and second-order equation system and introduce the notions of Lagrangian and mass transport velocity to discuss two potential boundary conditions for the second-order system. 
We reformulate the mass balance equation to provide both analytically and numerically consistent forms of mass source term for each of these boundary conditions.
Following Bradley~\etal~\cite{bradley1996acoustic}, we systematically relate these two velocities to clarify that a prescription for no-slip velocity at the oscillating boundary may not, in general, correspond to a zero mass transport velocity at the oscillating boundary. Subsequent to a formal verification of our numerical implementation, we consider both rectilinear and elliptical boundary oscillations to illustrate that prescription of a zero mass transport velocity at the oscillating boundary may lead to erroneous results, depending on the boundary oscillation profile. Lastly, we provide benchmark results for a test case with spatially varying fluid density and highlight the differences in streaming flow field compared to the homogeneous density case. Numerical solvers and benchmark test cases presented in this paper will facilitate a better understanding of multiphase acoustofluidic phenomena and will assist in benchmarking future numerical implementations involving multiphase/multicomponent fluids.

\section{Mathematical formulation} \label{sec_formulation}
\subsection{Continuous equations of motion}
The balance laws governing the motion of a compressible fluid in conservative form are given as
\begin{subequations} 
\begin{alignat}{2}
    &\frac{\partial \rho}{\partial t}+\bnabla \cdot(\rho \v)=0, \label{eq:mass} \\
    &\frac{\partial (\rho \v)}{\partial t}+\bnabla \cdot(\rho \v \otimes \v)=\bnabla \cdot \bm{\sigma}, \label{eq:momentum}
\end{alignat}
\end{subequations} 
in which $\rho$ is the mass density, $\v$ is the (Eulerian) fluid velocity, and $p$ is the fluid pressure. The Cauchy stress tensor $\bm{\sigma}$ for a linear, viscous compressible fluid can be expressed as
\begin{align}
    \bm{\sigma}=-p \V{I}+\mu(\nabla \v+(\nabla \v)^\intercal)+\lambda (\nabla \cdot \v) \V{I},
\end{align}
in which $\mu$ and $\lambda$ are spatially-varying shear and bulk viscosity, respectively. These equations need to be supplemented by an equation of state linking the fluid pressure to the mass density
\begin{align}
    p=\pfunc(\rho). \label{eq:state}
\end{align}
We employ the Nyborg's perturbation approach to linearize these equations by expanding the fluid velocity, density, and pressure as an infinite series 
\begin{subequations} 
\begin{alignat}{2}
& \v=\v_0+\epsilon \v'+\epsilon^2 \v''+\mathcal{O}(\epsilon^3),\\
& p=p_0+\epsilon p'+\epsilon^2 p''+\mathcal{O}(\epsilon^3),\\
& \rho=\rho_0+\epsilon \rho'+\epsilon^2\rho''+\mathcal{O}(\epsilon^3),
\end{alignat}
\end{subequations} 
in which $\epsilon$ is a non-dimensional smallness parameter defined as $\epsilon=d/a $, where $d$ is the displacement amplitude of the vibrating boundary and $a$ is a characteristic length. We introduce the notation $\v_1=\epsilon \v'$ and $\v_2=\epsilon^2 \v''$  (analogously for $p$ and $\rho$) such that the subscripts denote the order of the respective fields. The zeroth-order solution denotes the fluid state in the absence of acoustic actuation. In this work, we assume the fluid to be quiescent in the absence of acoustic actuation and set $\v_0$ to zero. The first-order fields denote the fluid's leading-order oscillatory response, while the (time-averaged) second-order fields denote the fluid's mean response. 

Substituting the perturbation expansion of the unknown fields into the governing equations (Eqs.~\eqref{eq:mass}--~\eqref{eq:state}), and gathering terms of $\mathcal{O}(\epsilon)$ leads to the following first-order system of equations (see~\ref{sec_sep_orders})
\begin{subequations} \label{eq:first-order-eqs}
\begin{alignat}{2}
&\frac{\partial \rho_1}{\partial t}+\bnabla \cdot\left(\rho_0 \v_1\right)=0,\label{eq:first-mass}\\
&\frac{\partial (\rho_0 \v_1)}{\partial t}=\bnabla \cdot \bm{\sigma}_1, \label{eq:first-momentum}\\
&\bm{\sigma}_1=-p_1 \V{I}+\mu(\bnabla \v_1+(\bnabla \v_1)^\intercal)+\lambda (\bnabla \cdot \v_1) \V{I},\label{eq:first-sigma}\\
&p_1 = c_0^2\rho_1, \label{eq:first-state}
\end{alignat}
\end{subequations} 
in which $c_0$ denotes the speed of sound in the fluid.

We consider the first-order fields to be periodic in time with time period $T$ and angular frequency $\omega = 2\pi/T$. We seek solutions of the form
\begin{subequations}  \label{eq:first-order-harmonic}
\begin{alignat}{2}
& \v_1(\r, t)  && =  \vonehat(\r)e^{\iota \omega t} = (\vone^{\rm r} (\r) + \iota \vone^{\rm i} (\r))e^{\iota \omega t}, \label{eq:v-harmonic} \\
& p_1(\r, t)  &&= \ponehat(\r)e^{\iota \omega t} = (\pone^{\rm r}  (\r)+ \iota \pone^{\rm i} (\r))  e^{\iota \omega t}, \label{eq:p-harmonic} \\
& \rho_1(\r, t) &&= \rhoonehat(\r)e^{\iota \omega t} = (\rhoone^{\rm r}  (\r)+ \iota \rhoone^{\rm i} (\r))  e^{\iota \omega t}, \label{eqn:rho-harmonic}
\end{alignat}
\end{subequations} 
in which the superscripts `$\rm r$' and `$\rm i$' denote the real (Re) and imaginary (Im) components of the spatially-varying part of a first-order quantity, respectively.  
%
Next, we repeat the same approach by gathering $\mathcal{O}(\epsilon^2)$ terms to obtain the governing equations at the second-order (see~\ref{sec_sep_orders}). They read as 
\begin{subequations}
\label{eq_second_order}
\begin{alignat}{2}
&\frac{\partial \rho_2}{\partial t}+ \bnabla \cdot (\rho_0 \v_2)=-\bnabla \cdot(\rho_1 \v_1),\\
&\frac{\partial (\rho_0\v_2)}{\partial t}+\frac{\partial (\rho_1 \v_1)}{\partial t}+\bnabla \cdot(\rho_0 \v_1 \otimes \v_1)=\bnabla \cdot \bm{\sigma}_2,\\
&\bm{\sigma}_2=-p_2 \V{I}+\mu(\bnabla \v_2+(\bnabla \v_2)^\intercal)+\lambda (\bnabla \cdot \v_2) \V{I},\\
&p_2 = c_0^2  \rho_2 + \frac{1}{2} \frac{\partial^2 p}{\partial \rho^2}\bigg\lvert_{\rho=\rho_0} \rho_1^2.
\end{alignat}
\end{subequations}
%
%
Given our interest in investigating steady acoustic streaming that is observed on large time scales compared to the acoustic actuation period, we neglect the time dependence of the second-order fields and 
apply a temporal averaging operation of the form $\langle A\rangle=\frac{1}{T} \int_T A\; \d t$. This time averaging operation yields the following system of equations
\begin{subequations}
\begin{alignat}{2}
&\bnabla \cdot \langle\rho_0 \v_2\rangle=-\bnabla \cdot\langle\rho_1 \v_1\rangle,\label{eq:second-mass}\\
&\bnabla \cdot (\langle\bm{\sigma}_2\rangle-\langle\rho_0\v_1 \otimes \v_1\rangle)=0,\\
&\langle\bm{\sigma}_2\rangle=-\langle p_2 \rangle \V{I}+\mu(\bnabla \langle \v_2 \rangle+(\bnabla \langle\v_2\rangle)^\intercal)+\lambda (\nabla \cdot \langle\v_2\rangle) \V{I},
\end{alignat}
\end{subequations}
in which we have used $\langle \frac{\partial (\rho_1 \v_1)}{\partial t} \rangle = 0$ since $\langle \partial_t (a b) \rangle =0$ for any two oscillating quantities $a$ and $b$ of time period $T$; see~\ref{proof:timederivative_product}.
We note that the equation of state (see~\ref{sec:EoS}) can be used to obtain second-order density via post-processing, if desired, but is not needed to solve for the second-order velocity and pressure.
Further, since the time-average of a zeroth- and second-order quantity is equal to the quantity itself, in the rest of this article, we drop the angle brackets around the terms containing zeroth- and second-order quantities and carry angle brackets only in terms which feature the product of two first-order quantities.

\subsection{Boundary conditions}
\label{sec: bc_cont}
In typical micro-acoustofluidic devices, the domain of interest corresponds to a liquid-filled microfluidic channel. The acoustic actuation of the device is generally modeled by prescribing a known time-periodic oscillatory motion to one or more boundaries of the domain. We denote these oscillating boundaries by $\Gamma^\textrm{p}$ and the remaining boundaries by $\Gamma^\textrm{w}$ such that the boundary of the microfluidic channel ($\Gamma$) is $\Gamma=\Gamma^\textrm{p} \cup \Gamma^\textrm{w}$. In this work, we consider the remaining boundaries of the domain ($\Gamma^\textrm{w}$) to be completely fixed. This scenario typically corresponds to the case where the microchannel wall is made of an \emph{acoustically-hard} material~\cite{bruus2012acoustofluidics}. Following Bradley~\cite{bradley1996acoustic}, let $\r_0$ denote the position of the oscillatory boundary at rest (i.e., in the absence of acoustic actuation). The oscillatory displacement of this boundary is prescribed in a Lagrangian sense as being $\presdisp (\r_0,t)$, such that the position of the deformed surface of the oscillatory boundary is given as $\r=\r_0 + \presdisp (\r_0,t)$. The no-slip boundary condition requires that the fluid and solid velocities at the boundary be equal
\begin{align}
    \v (\r,t) = \dot{\presdisp}(\r_0,t) \quad \textrm{for} \quad \r=\r_0 + \presdisp (\r_0,t), \label{eq:noslip}
\end{align}
in which $\dot{\presdisp}$ denotes the material derivative of the boundary displacement. Expanding the left hand side of Eq.~\eqref{eq:noslip} around the mean boundary position $\r_0$ using Taylor series expansion yields
\begin{align}
    \v (\r_0,t) +\bnabla \v (\r,t)\Bigr\rvert_{\r=\r_0}  \presdisp (\r_0,t) + \mathcal{O} (\epsilon^3) = \dot{\presdisp}(\r_0,t) \quad \textrm{for} \quad \r=\r_0 + \presdisp (\r_0,t).\label{eq: TS_expansion}
\end{align}
Perturbation expansion of the fluid velocity $\v=\v_1+\v_2+\mathcal{O} (\epsilon^3)$ on the left hand side of the above equation yields
\begin{align}
    \v_1 (\r_0,t) + \v_2 (\r_0,t)+ \bnabla \v_1 (\r,t)\Bigr\rvert_{\r=\r_0}  \presdisp (\r_0,t) + \mathcal{O} (\epsilon^3) = \dot{\presdisp}(\r_0,t) \quad \textrm{for} \quad \r=\r_0 + \presdisp (\r_0,t). \label{eq:TSexpansion}
\end{align}
in which $\presdisp$ is taken to be $\mathcal{O}(\epsilon)$. Matching orders of $\epsilon$ on the left and right hand sides of Eq.~\eqref{eq:TSexpansion} yields the first- and (time-averaged) second-order boundary conditions
\begin{subequations}
\begin{alignat}{2}
    &\v_1 (\r_0,t) = \dot{\presdisp}(\r_0,t) ,\\
    & \v_2 (\r_0,t)+ \langle\bnabla \v_1 (\r,t)\Bigr\rvert_{\r=\r_0}  \presdisp (\r_0,t) \rangle = \bm{0}.\label{eq:SDTS}
\end{alignat}
\end{subequations}
Note that in Eq.~\eqref{eq:SDTS} we have dropped angle brackets around $\v_2$ as it already assumed to be a time-averaged quantity. This also holds for $p_2$. The second term in Eq.~\eqref{eq:SDTS} is referred to as Stokes drift
\begin{equation}
    \v^\textrm{SD}= \langle \bnabla \v_1 (\r,t)\Bigr\rvert_{\r=\r_0}  \presdisp (\r_0,t) \rangle.\label{eq: SD}
\end{equation}
In the context of Generalized Lagrangian Mean theory~\cite{buhler2014waves}, the sum of the Eulerian velocity and Stokes drift is viewed as being the Lagrangian velocity of the fluid particle 
\begin{equation}
    \v^\textrm{L}:=\v_2 + \v^\textrm{SD}.\label{eq:vL}
\end{equation}
From a mathematical perspective, $\v^\textrm{SD}$ represent the second-order corrections to the fluid's Eulerian velocity that arise due to the oscillations of the boundary. As such, Eq.~\eqref{eq:SDTS} imposes the no-slip boundary condition, correct upto $\mathcal{O}(\epsilon^2)$,  at the displaced position of the oscillating boundary.
We note that the boundary condition in Eq.~\eqref{eq:SDTS} is obtained purely from kinematic consideration of enforcing no-slip at the oscillating wall without reference to the balance laws.

\hglt{Another boundary condition that has been employed in the acoustofluidic literature is based on the notion of mass transport velocity~\cite{das2022extra,muller2014numerical,muller2015theoretical}. In the rest of this section, we will clarify the relation between Lagrangian and mass transport velocities.}
To this end, we re-arrange the second-order mass balance in Eq.~\eqref{eq:second-mass} as
\begin{align}
   \nabla \cdot \left(\rho_0 \v^\textrm{M}\right) = 0, \quad \quad \textrm{with} \qquad \v^\textrm{M}:=\v_2 + \frac{1}{\rho_0 }\langle\rho_1 \v_1\rangle, \label{eq:vM} 
\end{align}
in which $\v^\textrm{M}$ is typically referred to as the mass transport velocity.
Combining Eqs.~\eqref{eq:vL} and \eqref{eq:vM}, one can relate $\v^\textrm{L}$ and $\v^\textrm{M}$ as
\begin{equation}
    \v^\textrm{L}=\v^\textrm{M}-\frac{1}{\rho_0 }\langle\rho_1 \v_1\rangle+ \v^\textrm{SD}.\label{eq:vLvMrel}
\end{equation}
With some algebraic manipulations (see \ref{proof:vl=vm+curl}), Eq.~\eqref{eq:vLvMrel} can be reformulated as
\begin{equation}
\v^\textrm{L}=\v^\textrm{M}-\frac{1}{2 \rho_0 }\langle\ \bnabla \times(\rho_0 \bm{\xi}_1 \times \vone)\rangle,\label{eq: vLvMrel_reform}
\end{equation}
in which  $\bm{\xi}_1 = \int \v_1 \d t$ is the first-order fluid displacement field. At the oscillating boundaries $\Gamma^{\rm p}$, $\bm{\xi}_1 \equiv \presdisp$ is prescribed. The last term on the right hand side of Eq.~\eqref{eq: vLvMrel_reform} depends on the boundary oscillation profile and therefore, can generally be non-zero. Accordingly,  $\v^\textrm{L}$ and $\v^\textrm{M}$ represent distinct notions of the mean velocity that become identical only under special circumstances; for example, when the microfluidic channel walls undergo rectilinear oscillations. Since the divergence of curl of any sufficiently differentiable vector field is zero, we have $\bnabla \cdot \langle\ \bnabla \times(\rho_0 \bm{\xi}_1 \times \vone)\rangle = 0$. This implies
\begin{equation}
\bnabla \cdot (\rho_0 \v^\textrm{L})=\bnabla \cdot (\rho_0 \v^\textrm{M}) = 0,\label{eq:vLvM_div}
\end{equation}
in which we have used Eq.~\eqref{eq:vM}. Combining Eq.~\eqref{eq:vLvM_div} with the definitions of $\v^\textrm{L}$ and $\v^\textrm{M}$ in Eqs.~\eqref{eq:vL} and \eqref{eq:vM} indicates that 
\begin{equation}
\bnabla \cdot \langle\rho_1 \v_1\rangle=\bnabla \cdot (\rho_0 \v^\textrm{SD}),\label{eq:vSD_vbolus}
\end{equation}
which, in turn, implies that the source term in the second-order mass balance equation (Eq.~\ref{eq:second-mass}) can be reformulated in terms of $\v^\textrm{SD}$. Nonetheless, while this reformulation is equivalent from an analytical standpoint, these forms of mass-source are not identical from a numerical perspective, as discussed later in Sec.~\ref{sec_matrix}. \hglt{Further, noting Eq.~\eqref{eq:vLvM_div}, while both $\v^\textrm{L} = \bm{0}$ and $\v^\textrm{M} = \bm{0}$ represent consistent choices of boundary conditions for the second-order system, the two boundary conditions are not identical since $\v^\textrm{L}$ and $\v^\textrm{M}$ are equal only when the last term in Eq.~\eqref{eq: vLvMrel_reform} vanishes.
As will be shown later through our numerical results in Section~\ref{subsec: elliptical}, depending on the oscillation profile, imposing the $\v^\textrm{M} = \bm{0}$ boundary condition can lead to physically incorrect solutions and may result in a non-zero Lagrangian mass transport at the oscillating boundary. }




\subsection{Discretized equations of motion}

\subsubsection{Spatial discretization} \label{sec_spatial_descritization}

\begin{figure}
    \centering
    \includegraphics[width=\linewidth]{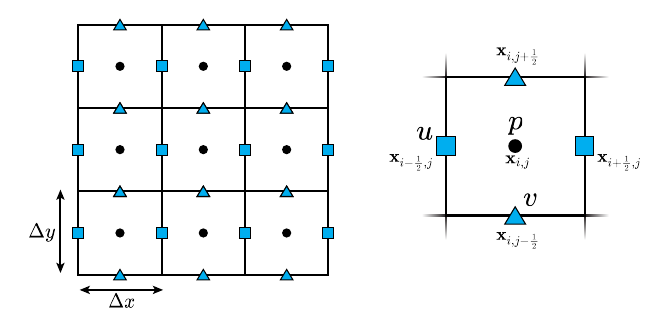}
    \caption{Schematic of a 2D staggered Cartesian grid illustrating (a) the coordinate system for the staggered grid; and (b) a single grid cell with (first- and second-order) velocity components $u$ ({\color{cyan} $\blacksquare$}) and $v$
({\color{cyan} $\blacktriangle$}) approximated at the cell faces  and scalar  pressure approximated at the cell center ({\color{black} $\bullet$}).}
    \label{fig_discretized_staggered_grid}
\end{figure}

The continuous equations of motion are discretized on a uniform staggered Cartesian grid. The computational domain $\Omega$ is divided into $N_x \times N_y$ rectangular cells. The cell size in the $x$ and $y$ directions is $\dx$ and $\dy$, respectively, as illustrated in Fig.~\ref{fig_discretized_staggered_grid}.
Unless stated otherwise, a uniform grid spacing  $\Delta x = \Delta y = \Delta$  is used for all simulations in this work.  We assume that the bottom-left corner of the computational domain $\Omega$ aligns with the origin $(0, 0)$. The position of each center of the grid cell is given by $\x_{i,j} = \left((i + \half)\dx,(j + \half)\dy\right)$, where $i = 0, \ldots, N_x - 1$ and $j = 0, \ldots, N_y - 1$. The face center in the $x-$direction, which is located half a grid space away from the cell center $\x_{i,j}$ in the negative $x-$direction, is represented by $\x_{i-\half,j} = \left(i\dx,(j + \half)\dy\right)$, where $i = 0, \ldots, N_x$ and $j = 0, \ldots, N_y - 1$. Similar conventions apply to other face center locations. First- and second-order pressure fields ($p_1$ and $p_2$, respectively) are stored at the cell centers. The $x-$component of first- and second-order velocities ($u_1$ and $u_2$, respectively) are stored at the centers of  $x-$direction cell faces, while the $y-$ component of first- and second-order velocities ($v_1$ and $v_2$, respectively) are stored at the centers of the faces of the $y-$ direction cell, as shown in Fig.~\ref{fig_discretized_staggered_grid}. Material properties, including density ($\rho$), shear viscosity ($\mu$) and bulk viscosity ($\lambda$) are stored at the cell centers. Second-order interpolation is used to interpolate the cell-centered quantities to faces and nodes, as required by the discrete spatial operators. Standard second-order finite differences are employed to approximate spatial differential operators. The spatial discretizations of the key continuous operators are as follows:
\begin{itemize}
\item The density-weighted divergence of the velocity field
$\v = (u,v)$ is approximated at cell centers by
\begin{subequations}
\begin{alignat}{2}
\label{eq_div_fd}
& \Drho(\v) = D^x_\rho \, u + D^y_\rho \, v, \\
&(D^x_\rho \, u)_{i,j} = \frac{\rho_{i+\half, j}u_{i+\half, j} - \rho_{i-\half, j} u_{i-\half, j}}{\dx}, \\
&(D^y_\rho \, v)_{i,j} = \frac{\rho_{i, j+\half} v_{i, j+\half} - \rho_{i, j-\half}v_{i, j-\half}}{\dy}. 
\end{alignat}
\end{subequations}
\item The gradient of cell-centered pressure is approximated at cell faces as
\begin{subequations}
\begin{alignat}{2}
\label{eq_grad_fd}
& \G p = (G^x p, G^y p), \\
&(G^x p)_{i-\half,j} = \frac{p_{i,j} - p_{i-1,j}}{\dx}, \\
&(G^y v)_{i,j - \half} =\frac{p_{i,j} - p_{i,j-1}}{\dy}. 
\end{alignat}
\end{subequations}
\item The continuous form of divergence of viscous strain rate tensor, which couples the velocity components through spatially variable shear viscosity, is given by
\begin{equation}
\label{eq_visc_cont}
\bnabla \cdot \left[\mu \left(\bnabla \v + (\bnabla \v)^\intercal\right) \right] = \left[
\begin{array}{c}
 (\Lmu \v)^x_{i-\half,j} \\
 (\Lmu \v)^y_{i,j-\half}  \\
\end{array}
\right] = 
\left[
\begin{array}{c}
 2 \D{}{x}\left(\mu \D{u}{x}\right) + \D{}{y}\left(\mu\D{u}{y}+\mu\D{v}{x}\right) \\
 2 \D{}{y}\left(\mu \D{v}{y}\right) + \D{}{x}\left(\mu\D{v}{x}+\mu\D{u}{y}\right) \\
\end{array}
\right].
\end{equation}
The shear viscosity operator is discretized using standard second-order, centered finite differences
\begin{subequations}
\begin{alignat}{2}
 (\Lmu \v)^x_{i-\half,j} &= \frac{2}{\dx}\left[\mu_{i,j}\frac{u_{i+\half,j} - u_{i-\half,j}}{\dx} -
					        \mu_{i-1,j}\frac{u_{i-\half,j} - u_{i-\3half,j}}{\dx}\right] \nonumber \\ 
                    &+ \frac{1}{\dy}\left[\mu_{i-\half, j+\half}\frac{u_{i-\half,j+1} - u_{i-\half,j}}{\dy} - 
					         \mu_{i-\half, j-\half}\frac{u_{i-\half,j} - u_{i-\half,j-1}}{\dy}\right] \nonumber \\
	            &+ \frac{1}{\dy}\left[\mu_{i-\half, j+\half}\frac{v_{i,j+\half} - v_{i-1,j+\half}}{\dx} - 
					         \mu_{i-\half, j-\half}\frac{v_{i,j-\half} - v_{i-1,j-\half}}{\dx}\right] \label{eq_viscx_fd} \\				         
 (\Lmu \v)^y_{i,j-\half} &= \frac{2}{\dy}\left[\mu_{i,j}\frac{v_{i,j+\half} - v_{i,j-\half}}{\dy} -
					        \mu_{i,j-1}\frac{v_{i,j-\half} - v_{i,j-\3half}}{\dy}\right] \nonumber \\ 
                    &+ \frac{1}{\dx}\left[\mu_{i+\half, j-\half}\frac{v_{i+1,j-\half} - v_{i,j-\half}}{\dx} - 
					         \mu_{i-\half, j-\half}\frac{v_{i,j-\half} - v_{i-1,j-\half}}{\dx}\right] \nonumber \\
	            &+ \frac{1}{\dx}\left[\mu_{i+\half, j-\half}\frac{u_{i+\half,j} - u_{i+\half,j-1}}{\dy} - 
					         \mu_{i-\half, j-\half}\frac{u_{i-\half,j} - u_{i-\half,j-1}}{\dy}\right] \label{eq_viscy_fd},
\end{alignat}
\end{subequations}
in which the shear viscosity is required at both cell centers and nodes of the staggered grid (i.e., $ \mu_{i\pm\half, j\pm\half}$).
Node-centered quantities are obtained via interpolation by 
arithmetically averaging the neighboring cell-centered quantities. 

\item The continuous form of divergence of bulk viscosity strain rate tensor, which couples the velocity components through spatially variable bulk viscosity, is given by
\begin{equation}
\label{eq_bulk_cont}
\div \left[\lambda (\div \v) \V{I} \right] = \left[
\begin{array}{c}
 (\Llamb \v)^x_{i-\half,j} \\
 (\Llamb \v)^y_{i,j-\half}  \\
\end{array}
\right] = 
\left[
\begin{array}{c}
 \D{}{x}\left(\lambda (\D{u}{x} + \D{v}{y}) \right)  \\
 \D{}{y}\left(\lambda (\D{u}{x} + \D{v}{y}) \right) \\
\end{array}
\right].
\end{equation}

The bulk viscosity operator is discretized using standard second-order, centered finite differences as
\begin{subequations}
\begin{alignat}{2}
 (\Llamb \v)^x_{i-\half,j} &=  \frac{1}{\dx}\left[\lambda_{i, j} \left(\frac{u_{i+\half,j} - u_{i-\half,j}}{\dx} + 
					         \frac{v_{i,j+\half} - v_{i,j-\half}}{\dy} \right) - \right. \nonumber \\ 
                             & \hspace{3em} \left. \lambda_{i-1, j} \left(\frac{u_{i-\half,j} - u_{i-\frac{3}{2},j}}{\dx} + 
					         \frac{v_{i-1,j+\half} - v_{i-1,j-\half}}{\dy} \right) \right] \label{eq_bulkx_fd} \\	         
 (\Llamb \v)^y_{i,j-\half} &= \frac{1}{\dy}\left[\lambda_{i, j} \left(\frac{u_{i+\half,j} - u_{i-\half,j}}{\dx} + 
					         \frac{v_{i,j+\half} - v_{i,j-\half}}{\dy} \right) - \right. \nonumber \\ 
                             & \hspace{3em} \left. \lambda_{i, j-1} \left(\frac{u_{i+\half,j-1} - u_{i-\half,j-1}}{\dx} + 
					         \frac{v_{i,j-\half} - v_{i,j-\frac{3}{2}}}{\dy} \right) \right] \label{eq_bulky_fd}.
\end{alignat}
\end{subequations}

\end{itemize}

\subsection{Matrix form of the equations} \label{sec_matrix}

Substituting the time periodic solution (Eqs.~\eqref{eq:first-order-harmonic}) into the first-order mass balance Eq.~\eqref{eq:first-mass} yields 
\begin{equation} \label{eq:first-mass-vec} 
 \iota \omega (\rhoone^{\rm r} + \iota \rhoone^{\rm i}) + \bnabla \cdot \left [\rho_0 (\vone^{\rm r} + \iota \vone^{\rm i}) \right] = 0.
\end{equation}
Substituting $\rhoone^{\rm r}=\pone^{\rm r}/c_0^2$ and $\rhoone^{\rm i}=\pone^{\rm i}/c_0^2$ into Eq.~\eqref{eq:first-mass-vec}, followed by separation of real and imaginary terms yields two first-order mass balance equations:
\begin{subequations} \label{eq:first-mass-comps}
\begin{alignat}{2}
    \textbf{Imaginary:}\quad \bigg(\frac{\omega}{c_0^2}\bigg) \pone^{\rm r} + \bnabla \cdot (\rho_0 \vone^{\rm i}) &= 0, \label{eq:im_cont_first}\\
    \textbf{Real:}\quad -\bigg(\frac{\omega}{c_0^2}\bigg) \pone^{\rm i} + \bnabla \cdot (\rho_0 \vone^{\rm r}) &= 0. \label{eq:re_cont_first}
\end{alignat}
\end{subequations}
Similarly, substituting the time periodic solution (Eqs.~\eqref{eq:first-order-harmonic}) into the first-order momentum Eq.~\eqref{eq:first-momentum} yields 
\begin{equation}
\begin{split}
    \iota \omega \rho_0 ( \vone^{\rm r} + \iota \vone^{\rm i} )  = - \bnabla(\pone^{\rm r} + \iota \pone^{\rm i}) + \bnabla \cdot [\mu (\bnabla ( \vone^{\rm r} + & \iota \vone^{\rm i} ) + (\bnabla( \vone^{\rm r} + \iota \vone^{\rm i} ))^\intercal) ]\\
    &+ \bnabla \left [\lambda \bnabla \cdot ( \vone^{\rm r} + \iota \vone^{\rm i} ) \right].    
\end{split}
\end{equation}
Separating the real and imaginary parts gives two first-order momentum equations:
\begin{subequations} \label{eq:first-momentum-comps}
\begin{alignat}{2}
    \textbf{Imaginary:}\quad  \omega \rho_0 \vone^{\rm r} &= -\bnabla \pone^{\rm i}+\bnabla \cdot[\mu(\bnabla \vone^{\rm i}+(\bnabla \vone^{\rm i})^\intercal)]+\nabla[\lambda \bnabla \cdot \vone^{\rm i}], \\
    \textbf{Real:}\quad -\omega \rho_0 \vone^{\rm i} &= -\bnabla \pone^{\rm r}+\bnabla \cdot[\mu(\nabla \vone^{\rm r}+(\bnabla \vone^{\rm r})^\intercal)]+\nabla[\lambda \bnabla \cdot \vone^{\rm r}].
\end{alignat}
\end{subequations}
Eqs.~\eqref{eq:first-momentum-comps} and~\eqref{eq:first-mass-comps}, when written in a matrix form, read as
\begin{equation} \label{eq:matrix-first}
\begin{bmatrix}
\omega \vrho_0 & \V{L_{\mu+\lambda}} & \V{0} & \G\\
\V{L_{\mu+\lambda}} & -\omega \vrho_0 & \G & \V{0}\\
\V{0} & \Drho_0 & \frac{\omega}{c_0^2}\V{I} & \V{0}\\
\Drho_0 & \V{0} & \V{0} & -\frac{\omega}{c_0^2}\V{I}
\end{bmatrix}  
\begin{bmatrix}
\vone^{\rm r} \\
\vone^{\rm i} \\
\vpone^{\rm r} \\
\vpone^{\rm i}
\end{bmatrix} 
=\V{0},
\end{equation}
in which $\vrho_0$ denotes the zeroth-order density field, interpolated to the face centers. The discrete versions of the spatial operators in Eq.~\eqref{eq:matrix-first} are defined as
\begin{subequations} 
\begin{alignat}{2}
    & \Drho_0(\v)=\bnabla \cdot (\rho_0 \v),\\
    & \V{L_{\mu+\lambda}}(\v) = -\V{L_\mu} (\v) -\V{L_\lambda} (\v) =-\bnabla \cdot [\mu (\bnabla \v+(\bnabla \v)^\intercal)+\lambda (\bnabla \cdot \v) \V{I})],\\
    & \V{G}(\V{p})= \bnabla p.
\end{alignat}
\end{subequations}
At the second-order, we have a steady-state low-Mach Stokes system, which in matrix form $\V{A} \x = \b$ reads as 
\begin{align} \label{eq:matrix-second}
\begin{bmatrix}
\V{L_{\mu+\lambda}} & \G\\
-\Drho_0 & \V{0}
\end{bmatrix}  
\begin{bmatrix}
\v_2 \\
\vptwo
\end{bmatrix} 
&=
\begin{bmatrix}
-\langle \Drho_0(\v_1 \otimes \v_1)\rangle \\
\V{\mathcal{M}}
\end{bmatrix}.
\end{align}
Here, $\V{\mathcal{M}} = \Drho_0 (\v^\textrm{SD})$ when $\v_2 = -\v^{\rm SD}$ (or equivalently $\v^{\rm L}  = \V{0} $) is imposed as the boundary condition, and  $\V{\mathcal{M}} = \langle \Drho_1 (\v_1) \rangle$ when $\v_2 = -\frac{1}{\rho_0}\langle \rho_1 \vone\rangle$  (or equivalently $\v^{\rm M}  = \V{0} $) is used as the boundary condition. At the continuous level, both forms of $\V{\mathcal{M}}$ are equivalent; see Eq.~\eqref{eq:vSD_vbolus}. However, discretely, they differ. In our numerical simulations we observe that the iterative solver's convergence rate degrades severely if we interchange the form of $\V{\mathcal{M}}$ with the chosen boundary condition ($\v^{\rm L} = \V{0}$ or $\v^{\rm M} = \V{0}$). 

Many studies have used an inconsistent combination of $\V{\mathcal{M}}$ and boundary conditions for the second-order system. For example, refs.~\cite{muller2012numerical,nama2015numerical,das2019acoustothermal} used $\V{\mathcal{M}} = \langle \Drho_1 (\v_1) \rangle$ and $\v_2 = \V{0}$ as the boundary condition. This is generally an inconsistent choice because the boundary conditions for $\v_2$ should be such that the condition 
\begin{subequations} \label{eq:inconsistent_v2_bc}
\begin{alignat}{2}
& -\int_\Omega \bnabla \cdot \rho_0 \v_2 \; \d V =   \int_\Omega \V{\mathcal{M}} \; \d V, \\
\hookrightarrow & -\int_{\partial\Omega}  \rho_0 \v_2 \cdot \V{n}\; \d S = \int_{\partial\Omega}  \langle \rho_1 \v_1 \rangle \cdot \V{n}\; \d S,
\end{alignat}
\end{subequations}
must be satisfied. With $\v_2 = \V{0}$ as the imposed boundary condition for the second-order system, inconsistency arises if $\int_\Omega \V{\mathcal{M}} \; \d V$ does not vanish, which is typically the case.

\subsection{Linear solvers} \label{sec_linear_solvers}

The first-order system of equations (Eq.~\eqref{eq:matrix-first}) are essentially coupled Helmholtz equations in the complex amplitude of $\v_1$. This can be observed from Eqs.~\eqref{eq:first-mass-comps}, which  relates complex amplitudes of $p_1$ and $\v_1$. In the absence of a good iterative solver for the coupled Helmholtz equations, we utilize MUMPS~\cite{MUMPS:1}, which is a sparse direct solver to solve first-order equations to compute $\vonehat$ and $\ponehat$. In contrast, the second-order system of equations represent a steady-state, low-Mach Stokes system, for which a number of iterative solvers and preconditioners have been proposed in the literature. Following our success in employing a projection method-based preconditioner to solve multiphase incompressible Navier-Stokes equations~\cite{nangia2019robust}, we here combine the projection preconditioner with the flexible GMRES (FGMRES) solver to solve the steady-state low-Mach Stokes system (Eqs.~\eqref{eq:matrix-second}) for $\v_2$ and $p_2$. 

The projection method is typically used to solve time-dependent Navier-Stokes equations in the literature. Here, we describe a projection method for solving steady-state Stokes equations. We remark that we use projection method as a preconditioner and not as a solver. The projection method relies on the time-derivative term to split velocity and pressure solutions. The error incurred due to velocity-pressure splitting in time dependent systems is $\mathcal{O}(\Delta t)$ or $\mathcal{O}(\Delta t^2)$, depending upon the type of the projection algorithm employed~\cite{brown2001accurate}. It is necessary to solve velocity and pressure fields simultaneously for steady Stokes systems without a time derivative term. While the projection method cannot solve a steady Stokes system directly, it can speed up the convergence of the Krylov solver used to solve Eq.~\eqref{eq:matrix-second}. As a preconditioner, the projection method improves the current iterate of the outer Krylov solver by solving a residual equation of the form:
\begin{align}
\left[
\begin{array}{cc}
 \V{L_{\mu+\lambda}} & \G\\
 -\Drho_0 & \mathbf{0} \\
\end{array}
\right]
\left[
\begin{array}{c}
  \ev\\
  \ep \\
\end{array}
\right] & =
\left[
\begin{array}{c}
 \rv\\
 \rp \\
\end{array}
\right].
\end{align}
Here, $\ev$ and $\ep$ denote errors in the second-order velocity and pressure degrees of freedom, respectively, and the right-hand side vectors $\rv$ and $\rp$ are the residuals of the second-order momentum and mass-balance equations, respectively. 

In the first step of the projection method, an intermediate approximation to (second-order) velocity is computed by solving
\begin{equation}
\label{eq_frac_vel}
\V{L_{\mu+\lambda}} \; \widetilde{\e}_{\v} = \bv.
\end{equation}
The approximation $\widetilde{\e}_{\v}$ generally does not satisfy the discrete continuity equation,
i.e., $-\Drho_0 (\widetilde{\e}_{\v}) \ne \rp$. This condition can be satisfied by introducing an auxiliary
scalar field $\vvarphi$ and carrying out an operator splitting of the form 
\begin{subequations} 
\begin{alignat}{2}
(\ev - \widetilde{\e}_{\v}) &= -\G \vvarphi, \label{eq_frac_timestep} \\
 -\Drho_0(\ev) &= \rp \label{eq_frac_continuity}.
\end{alignat}
\end{subequations}
Taking the density-weighted divergence of Eq.~\eqref{eq_frac_timestep}, and making use of Eq.~\eqref{eq_frac_continuity} we obtain a variable-coefficient (due to $\rho_0$) Poisson equation for the scalar $\vvarphi$:
\begin{equation}
\label{eq:poisson_varphi}
-(\Drho_0 \G) \vvarphi = -\rp - \Drho_0(\widetilde{\e}_{\v}).
\end{equation}
The updated velocity solution can be computed from the solution of $\vvarphi$ as
\begin{equation}
\label{eq_frac_up_vel}
\ev = \widetilde{\e}_{\v} -  \G \vvarphi,
\end{equation}
and that of pressure can be computed as
\begin{equation}
\label{eq_frac_up_pressure}
\ep = \vvarphi.
\end{equation}
A more accurate estimate of pressure based on the spectral analysis of the Schur complement of Eq.~\eqref{eq:matrix-second} is given by~\cite{thirumalaisamy2023pre} 
\begin{equation}
\label{eq:pressure_schur}
\ep = -2 \V{\mu}(\rp + \Drho_0 (\widetilde{\e}_{\v})).
\end{equation}
Here, $\vvarphi$ and $\V{\mu}$ are both cell-centered quantities. In this work we approximate pressure through Eq.~\eqref{eq:pressure_schur} within the projection preconditioner. 

In our solution algorithm for solving the second-order system, we set a tight relative tolerance for the outer FGMRES solver. The outer Krylov solver (FGMRES) generates a Krylov subspace by applying the action of matrix $\A$ on vectors. It also requires the action of the projection preconditioner on residual vectors to get estimates on velocity and pressure errors. The FGMRES solver is deemed to be converged  if a value of $10^{-9}$ or below is reached for the norm of the relative residual $\mathcal{R} = \frac{||\r||}{||\b||} = \frac{||\b - \A \x||}{||\b||}$. In the projection preconditioner, we solve Eqs.~\eqref{eq_frac_vel} and~\eqref{eq:poisson_varphi} for velocity and pressure in an inexact manner. Specifically, we solve the velocity and pressure subdomain problems using a single iteration of Richardson solver that is preconditioned with a single V-cycle of a geometric multigrid solver. For both velocity and pressure problems, 3 iterations of Gauss-Seidel smoothing are performed on each multigrid level.

\subsection{\hglt{Software implementation}}
     
The perturbation-based variable-coefficient acoustofluidic solver described here is implemented within the IBAMR library~\cite{IBAMR-web-page}, an open-source C++ software enabling computational fluid dynamics (CFD) algorithm development. The code is hosted on GitHub at \url{https://github.com/IBAMR/IBAMR}.
IBAMR relies on SAMRAI \cite{HornungKohn02, samrai-web-page} for Cartesian grid 
management. Solver support in IBAMR is provided by the 
PETSc library~\cite{petsc-user-ref, petsc-web-page}.


\section{Results} \label{sec_results}

\subsection{Solver accuracy and convergence}
\label{accuracy and convergence}
In order to test the accuracy and convergence of our implementation of the coupled first-and second-order systems, we use the method of manufactured solution~\cite{salari2000code}. A computer code provides the numerical solution to some equations given a set of problem data that includes problem parameters, initial conditions, boundary conditions, and source terms. A numerical implementation must therefore demonstrate that it produces the expected output for a prescribed set of data in order to be verified. The method of manufactured solution comprises of the following sub-steps: (1) In Step 1, the solution of the problem is manufactured by assuming analytical expressions for each unknown. (2) This manufactured solution is used in Step 2 to determine the data to be provided to the numerical code to solve the problem. The governing equations are substituted with analytical expressions to obtain the source terms that are consistent with the manufactured solution. Additionally, the manufactured solution is evaluated at the boundary in order to generate boundary conditions. (3) Finally, the data generated in Step 2 (source terms and boundary conditions) are fed into the code to obtain a numerical solution that should ideally recover the manufactured solution from Step 1. Comparing the manufactured and numerical solution over multiple computational grids can help determine the method's convergence rate. 

The first-order problem given by Eq.~\eqref{eq:matrix-first} does not have any prescribed source terms. To apply the method of manufactured solution, we rewrite this problem by adding the requisite source terms in the momentum equation
\begin{equation} \label{eq:matrix-first-MS}
\begin{bmatrix}
\omega \vrho_0 & \V{L_{\mu+\lambda}} & \V{0} & \G\\
\V{L_{\mu+\lambda}} & -\omega \vrho_0 & \G & \V{0}\\
\V{0} & \Drho_0 & \frac{\omega}{c_0^2}\V{I} & \V{0}\\
\Drho_0 & \V{0} & \V{0} & -\frac{\omega}{c_0^2}\V{I}
\end{bmatrix}  
\begin{bmatrix}
\vone^{\rm r} \\
\vone^{\rm i} \\
\vpone^{\rm r} \\
\vpone^{\rm i}
\end{bmatrix} 
=
\begin{bmatrix}
\vec{f}_{\v}^{\,\rm r} \\
\vec{f}_{\v}^{\,\rm i} \\
\V{0} \\
\V{0}
\end{bmatrix} .
\end{equation}
The first-order mass balance equations do not include a mass source term in Eq.~\eqref{eq:matrix-first-MS}. This is because, in this test, we impose $\v^{\rm L} = \V{0}$ as the boundary condition for the second-order system and a consistent combination of $\V{\mathcal{M}} = \Drho_0(\v^{\rm SD})$ and $\v^{\rm L} = \V{0}$ as a boundary condition for the second-order system requires no mass source terms for the first-order system; see the derivation in~\ref{proof:vl=vm+curl}.

The computational domain is chosen to be a unit square: $\Omega = [0,L]^2$ with  $L=1\,\textrm{m}$ being the side length. We consider the fluid material parameters to be spatially varying by prescribing the mass density, shear viscosity, and bulk viscosity of the form
\begin{subequations}
\label{eq:fluid_parameters}
    \begin{alignat}{2}
        \rho_0 &= A + B x^2y,\\
        \mu &= C + D x^2y,\\
        \lambda &= C + D x^2y,
    \end{alignat}
\end{subequations}
respectively, in which $A$, $B$, $C$, and $D$ are constants whose values are taken as $10\,\si{kg \cdot m^{-3}}$, $1\,\si{kg \cdot m^{-6}}$, $10\,\si{Pa \cdot s}$, and $1\,\si{Pa \cdot s}$,  respectively. The angular frequency is taken to be $\omega=1\,\textrm{Hz}$. The manufactured solution for the first-order velocity is chosen as 
\begin{subequations} \label{eq:manu-v1}
    \begin{alignat}{2}
    \vone^{\rm r}=& (x^3+y^3, x^2+y^2),\\
    \vone^{\rm i}=& (x^2+y^2, x^3+y^3),
    \end{alignat}
\end{subequations}
in which $c_0$ is the speed of sound in the fluid which is taken to be $1\,\si{m \cdot s^{-1}}$. This manufactured velocity $\v_1$ is then substituted in Eq.~\eqref{eq:matrix-first-MS} to determine first-order pressure fields, which read as 
\begin{subequations}
    \begin{alignat}{2}
    \pone^{\rm r}=& \frac{c_0^2}{\omega}(x^5+4x^3y+4x^2y^3+30 y^2 + 2x (y^3 + 10)),\\
    \pone^{\rm i}=& \frac{c_0^2}{\omega}(2xy^4 +x^4+5x^4y + 20 y +3x^2(y^2+ 10)).
    \end{alignat}
\end{subequations}
The first-order velocity and pressure fields are then substituted in Eq.~\eqref{eq:matrix-first-MS} to determine the momentum source terms $\vec{f}_{\v}^{\,\rm r}$ and $\vec{f}_{\v}^{\,\rm i}$. These source terms are provided as input for our numerical implementation. Additionally, we prescribe Dirichlet boundary conditions for $\vone^{\rm r}$ and $\vone^{\rm i}$ on the entire domain boundary by using the corresponding manufactured solution. The numerical solution of Eq.~\eqref{eq:matrix-first-MS} is compared against the manufactured solution for $\v_1^{\rm r},\v_1^{\rm i},p_1^{\rm r},\text{ and } p_1^{\rm i}$. The $L^1$ and $L^2$ norms of error $\V{\mathcal{E}}$ between the numerical  and manufactured solutions for cell-centered pressure $p$ are calculated as 
\begin{subequations}
\begin{alignat}{2}
    \hglt{||\V{\mathcal{E}}_p||_{L^1}}=  &\sum_{i=0}^{N_x-1}\sum_{j=0}^{N_y-1} |\mathcal{E}_{p_{i,j}}|\Delta x \Delta y,\\
    \hglt{||\V{\mathcal{E}}_p||_{L^2}} = &\left(\sum_{i=0}^{N_x-1}\sum_{j=0}^{N_y-1} \mathcal{E}_{p_{i,j}}^2\Delta x \Delta y \right)^{\frac{1}{2}}.
\end{alignat}
\end{subequations}
For side-centered velocity $\v$, the $L^\beta$ norm ($\beta=1,2$) is computed as $\lVert\v\rVert^\beta = \lVert u \rVert^\beta + \lVert v \rVert^\beta$ with
\begin{subequations}
\begin{alignat}{4}
    \hglt{||\V{\mathcal{E}}_u||_{L^1}} =  &\frac{1}{2}\sum_{j=0}^{N_y-1} \left|\mathcal{E}_{u_{-\frac{1}{2},j}}\right| \Delta x \Delta y+\sum_{i=0}^{N_x-1}\sum_{j=0}^{N_y-1} \left|\mathcal{E}_{u_{i-\frac{1}{2},j}}\right|\Delta x \Delta y + \frac{1}{2}\sum_{j=0}^{N_y-1} \left|\mathcal{E}_{u_{N_x-\frac{1}{2},j}}\right| \Delta x \Delta y, \\
    \hglt{||\V{\mathcal{E}}_v||_{L^1}} =  &\frac{1}{2}\sum_{i=0}^{N_x-1} \left|\mathcal{E}_{v_{i,-\frac{1}{2}}}\right| \Delta x \Delta y+\sum_{i=0}^{N_x-1}\sum_{j=0}^{N_y-1} \left|\mathcal{E}_{v_{i,j-\frac{1}{2}}}\right|\Delta x \Delta y + \frac{1}{2}\sum_{i=0}^{N_x-1} \left|\mathcal{E}_{v_{i,N_y-\frac{1}{2}}}\right| \Delta x \Delta y, \\
    \hglt{||\V{\mathcal{E}}_u||_{L^2}} =  &\left(\frac{1}{2}\sum_{j=0}^{N_y-1} \mathcal{E}_{u_{-\frac{1}{2},j}}^2 \Delta x \Delta y \right)^{1/2}+\left(\sum_{i=0}^{N_x-1}\sum_{j=0}^{N_y-1} \mathcal{E}_{u_{i-\frac{1}{2},j}}^2\Delta x \Delta y\right)^{1/2} + \left(\frac{1}{2}\sum_{j=0}^{N_y-1} \mathcal{E}_{u_{N_x-\frac{1}{2},j}}^2 \Delta x \Delta y\right)^{1/2}, \\
    \hglt{||\V{\mathcal{E}}_v||_{L^2}} =  &\left(\frac{1}{2}\sum_{i=0}^{N_x-1} \mathcal{E}_{v_{i,-\frac{1}{2}}}^2 \Delta x \Delta y \right)^{1/2}+\left(\sum_{i=0}^{N_x-1}\sum_{j=0}^{N_y-1} \mathcal{E}_{v_{i,j-\frac{1}{2}}}^2\Delta x \Delta y\right)^{1/2} + \left(\frac{1}{2}\sum_{i=0}^{N_x-1} \mathcal{E}_{v_{i,N_y-\frac{1}{2}}}^2 \Delta x \Delta y\right)^{1/2}.
\end{alignat}
\end{subequations}

\begin{figure}
    \centering
    \includegraphics[width=\linewidth]{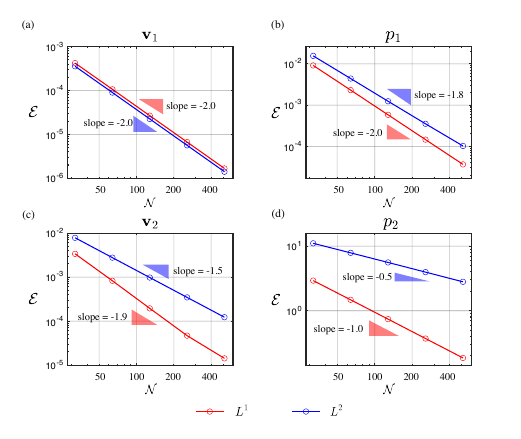}
    \caption{Spatial convergence rate of errors for the method of manufactured solution. Each panel presents a log-log plot of $L^1$ (red curve) and $L^2$ (blue curve) error norms as a function of number of grid cells ($\mathcal{N}$) in each direction of the 2D $\mathcal{N}\times \mathcal{N}$ square domain. Figs. (a) and (b) plot the error norms for first-order velocity and pressure, respectively. Figs. (c) and (d) plot the error norms for second-order velocity and pressure, respectively. The slope in each figure represents the convergence order.}
    \label{fig:convergence}
\end{figure}

Figs.~\ref{fig:convergence}(a,b) plot the $L^1$ and $L^2$ norm of errors in velocity and pressure as a function of $\mathcal{N}$, in which $\mathcal{N}$ represents the number of grid cells along each direction of the square domain. 
Near second-order convergence is observed for both velocity and pressure in both norms, as 
expected for our second-order accurate finite differencing and interpolation schemes.

Moving to the second-order problem, we modify Eq.~\eqref{eq:matrix-second} by adding the source terms as
\begin{align} \label{eq:matrix-second-MS}
\begin{bmatrix}
\V{L_{\mu+\lambda}} & \G\\
-\Drho_0 & \V{0}
\end{bmatrix}  
\begin{bmatrix}
\v_2 \\
\vptwo
\end{bmatrix} 
&=
\begin{bmatrix}
-\langle \Drho_0(\v_1 \otimes \v_1)\rangle \\
\Drho_0(\v^{\rm SD})
\end{bmatrix}
+
\begin{bmatrix}
\vec{s}_{\v} \\
\vec{s_{p}} = \V{0}
\end{bmatrix}.
\end{align}
A special manufactured solution for $\v_2$ is needed to ensure that the mass source term $\V{s_p} = \V0$. For proper convergence of the iterative solver for a low-Mach system, $\V{s_p} = \V{0}$ must be set. If $\V{s_p}$ is non-zero, it should be derived from the problem's physics and computed numerically. There is a possibility that the iterative solver will diverge when it encounters a non-physical mass source term $\V{s_p}$ (e.g., from an analytically generated manufactured solution). If we take the second-order velocity of the form $\v_2 = -\v^{\rm SD} + \v^\textrm{ div-free}$, then this requirement is naturally satisfied. We also need to impose the boundary condition $\v^{\rm L} = \V{0}$ or $\v_2 = -\v^{\rm SD}$ for the second-order system. Thus, we manufacture $\v_2 = -\v^{\rm SD}$, which in turn can be generated entirely from $\v_1$ as
\begin{align}
\v_2 = -\langle \bnabla \v_1 \V{\xi}_1\rangle = -\frac{1}{2}\operatorname{Re}\left(\bnabla \vonehat \widehat{\V{\xi}_1}^*\right) = -\frac{1}{2}\operatorname{Re}\left(\bnabla \vonehat \left\{\frac{\vonehat}{\iota \omega}\right\}^*\right),\label{eq:manu_v2}
\end{align}
in which $(\bullet)^*$ denotes the complex conjugate the quantity $(\bullet)$. Here, we have used Eq.~\eqref{eq:<ab>} to express the time-average of two oscillating quantities in terms of their complex amplitudes; see~\ref{sec_<ab>} for proof. Substituting Eqs.~\eqref{eq:manu-v1} into Eq.~\eqref{eq:manu_v2} yields the components of $\v_2$
\begin{subequations}
    \begin{alignat}{2}
        u_2 =& \frac{-1}{2\omega}((3x^2-2y)(x^2+y^2)+(3y^2-2x)(x^3+y^3)),\\
        v_2 = & \frac{-1}{2\omega}((2x-3y^2)(x^2+y^2)+(2y-3x^2)(x^3+y^3)).
    \end{alignat}
\end{subequations}
The manufactured solution for second-order pressure is taken to be
\begin{equation}
p_2 = xy + x^2y^2.
\end{equation}
Substituting the manufactured solutions for both the first- and second-order variables in Eq.~\eqref{eq:matrix-second-MS}, gives us an analytical expression (not shown here for brevity) for the forcing term $\vec{s}_{\v}$, which is fed to the numerical solver as the data. 

Figs.~\ref{fig:convergence}(c,d) plot the $L^1$ and $L^2$ norm of errors in second-order velocity and pressure for a series of computational grids. For velocity, the order of accuracy is between 1.5 and 2, whereas for pressure, it is between 0.5 and 1. The reduction of accuracy for the second-order system is attributed to the numerical evaluation of source terms $-\langle \Drho_0(\v_1 \otimes \v_1)\rangle$ and $\Drho_0(\v^{\rm SD})$, as well as numerically computing $\v^{\rm SD}$ to impose Dirichlet boundary conditions. In these evaluations, second-order accurate finite difference stencils are applied to the discrete solution $\v_1$, which is itself only second-order accurate. Similar reduction in accuracy has also been alluded by Muller~\etal~\cite{muller2014numerical} in the context of finite element analysis of acoustofluidic problems. Nonetheless, we also solved a decoupled second-order system that did not require a first-order solution. In that case, we obtained second-order accurate solutions for $\v_2$ and $p_2$ illustrating the expected accuracy of our numerical implementation; see \ref{accuracy and convergence_decoupled}.

\subsection{Impact of boundary oscillation profile}
Having verified our numerical implementation via manufactured solution, we now present numerical results for different boundary oscillation profiles and discuss the associated boundary conditions. Specifically, we focus on two commonly employed boundary oscillation profiles: (i) rectilinear; and (ii) elliptical oscillations. The rectilinear oscillation profile is commonly used in resonant bulk acoustic wave devices where a standing acoustic wave is often setup within a resonant microfluidic cavity~\cite{muller2012numerical}. In contrast, the elliptical oscillation profile is usually used to model surface acoustic wave devices which establish a surface wave on a piezoelectric substrate~\cite{barnkob2018acoustically}. While these oscillation profiles have been employed in several modeling studies of acoustofluidic devices, the handling of second-order mass source terms and the associated boundary conditions for the second-order problem remains a point of confusion.
With regards to the second-order boundary condition, different numerical studies have suggested homogeneous Dirichlet boundary conditions for different notions of mean velocity. Some studies prescribe homogeneous Dirichlet boundary conditions for Eulerian fluid velocity ($\v_2$)~\cite{nama2015numerical,muller2012numerical,das2019acoustothermal,ghorbani2022acoustic} while others prescribe zero Lagrangian velocity ($\v^\textrm{L}$)~\cite{kshetri2023acoustophoresis,kshetri2024evaluating,baasch2019acoustic,pavlic2022influence} or zero mass transport velocity ($\v^\textrm{M}$)~\cite{das2022extra,muller2014numerical,muller2015theoretical}. As discussed earlier in Sec.~\ref{sec: bc_cont}, $\v_2$, $\v^\textrm{L}$, and $\v^\textrm{M}$ are distinct representations of the fluid's mean velocity that are not, in general, identical. 

Some prior studies~\cite{kshetri2023acoustophoresis,kshetri2024evaluating,baasch2019acoustic,pavlic2022influence,muller2012numerical} have retained the mass source in the second-order mass balance Eq.~\eqref{eq:second-mass} while others~\cite{lei2017transducer,sankaranarayanan2008flow,devendran2014separation} have chosen to omit this term completely. Referring to Eqs.~\eqref{eq:second-mass} and~\eqref{eq:vSD_vbolus}, we note that the mass source depends on the oscillating boundary velocity. The latter is prescribed as an input (boundary condition) to the first-order problem. Therefore, from a mathematical perspective, a homogeneous Dirichlet boundary condition on Eulerian velocity may, in general, be inconsistent with the presence of a mass source term; see for example Eqs.~\eqref{eq:inconsistent_v2_bc}. Among the reasons for the differences in handling the mass source term and boundary conditions, and the resulting confusion, is that, depending on the boundary oscillation profile and physical parameters, these different mean velocities can converge in such a way that the numerical results are essentially the same. Nevertheless, when dealing with general scenarios, boundary conditions need to be assigned with prudence. In light of this, we present numerical results for rectilinear and elliptical oscillation profiles to illustrate the discrepancies between different boundary conditions.  

\subsubsection{Rectilinear actuation}
\label{sec_rectilinear}
We consider a bulk acoustic wave driven acoustofluidic setup, as reported by Muller~\etal~\cite{muller2012numerical}. The computational domain consists of a rectangular two-dimensional fluid-filled microchannel with dimensions $W=\SI{380}{\micro\metre}$ and $H= \SI{160}{~\micro\metre}$ situated in the $x-y$ plane. The fluid within the microchannel is water with a uniform mass density $\rho_0=998~\si{kg\cdot m^{-3}}$ and viscosities $\mu=0.89~\si{mPa\cdot s}$, $\lambda=1.88~\si{mPa\cdot s}$. The left and right boundaries of the domain are harmonically actuated with a rectilinear displacement of the form
\begin{equation}
    \presdisp(x,t) = d_0 e^{\iota \omega t} \vec{e}_x\qquad \textrm{for} \quad x = 0, W \qquad \textrm{and} \quad 0 \leq y\leq H, \label{eq:BAWoscillation}
\end{equation}
in which $d_0 = 0.1\,\si{nm}$ is the displacement amplitude, $\omega = 2\pi f$ is the angular frequency, and $\vec{e}_x$ is a unit vector along the $x$ direction with the origin located at the bottom left corner of the microchannel. The top and bottom walls of the microchannel are assumed to be fixed. Based on the microchannel dimensions and the fluid considered, $f=1.97\,\si{M Hz}$ represents a half-wave resonance mode and is taken as the actuation frequency in this case. We use $\V{\mathcal{M}} = \Drho_0 (\v^\textrm{SD})$ and $\v_2 = -\v^{\rm SD}$ as the boundary condition for the second-order system.

\begin{figure}
    \centering
    \includegraphics[width=0.80\linewidth]{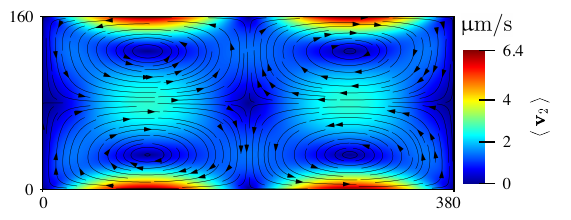}
    \caption{Streamlines and magnitude of the Eulerian streaming velocity $\langle\v_2\rangle$ solution obtained for the boundary oscillations corresponding to Eq.~\eqref{eq:BAWoscillation} and $\v^\textrm{L}=\vec{0}$ boundary condition for the second-order problem.}
    \label{fig:rectilinear}
\end{figure}

Fig.~\ref{fig:rectilinear}
plots the magnitude and streamlines of the Eulerian streaming velocity field obtained from our numerical implementation. Our results indicate the presence of four vortices inside the microchannel and are in excellent qualitative and quantitative agreement with those reported by Muller~\etal~\cite{muller2012numerical}. We note that in microfluidic experiments, the velocity field is typically obtained by tracking tracer microparticle trajectories. As such, a comparison of numerical and experimental results requires determining the particle velocity field. This corresponds to the fluid's Lagrangian velocity for a tracer particle. Referring to Eq.~\eqref{eq:vL}, the fluid's Lagrangian velocity can be computed via simple post-processing of the numerical result, once the Eulerian velocity field ($\vec{v}_2$) and Stokes drift ($\vec{v}^\textrm{SD}$) have been determined. In light of this, we remark that the Eulerian streaming velocity field obtained by Muller~\etal~\cite{muller2012numerical} (and reproduced in Fig.~\ref{fig:rectilinear} via our computational framework) does not incorporate the Stokes drift correction. Nonetheless, the Eulerian streaming velocity field has been shown to match experimentally-observed particle trajectories~\cite{muller2013ultrasound}. This finding seems to indicate that the Stokes drift correction, for the specific device and operational parameters considered here, may be negligible. This is true specifically in the bulk of the microfluidic channel where the trajectories were obtained experimentally and is indeed confirmed by our numerical results. The Lagrangian velocity field $(\vec{v}^\textrm{L})$ is virtually identical to the Eulerian velocity field in the bulk of the microfluidic channel. 

The boundary condition in Eq.~\eqref{eq:SDTS} prescribes a non-zero velocity field at the oscillating boundary. In contrast, Muller~\etal~\cite{muller2012numerical} prescribe a homogeneous Dirichlet boundary condition on the Eulerian velocity field ($\vec{v}_2=\vec{0}$). For the specific oscillation profile in Eq.~\eqref{eq:BAWoscillation}, the Stokes drift has a non-zero transverse component at the oscillating walls. Therefore, the Eulerian and Lagrangian velocity at the oscillating walls are not identical. As such, the prescription of zero Eulerian velocity on the oscillating walls is inconsistent with the presence of a mass source in Eq.~\eqref{eq:second-mass} and results in the convection of a mass source across the oscillating boundary. From a numerical perspective, the prescription of boundary conditions inconsistent with the governing equations may lead to non-convergence of the linear solver. However, for commonly considered high-frequency acoustofluidic setups, the mass source term in Eq.~\eqref{eq:second-mass} is small and can typically be neglected~\cite{baasch2019acoustic}. In such scenarios, the numerical implementation converges to a solution even when a zero Eulerian velocity boundary condition, which violates the second-order mass balance, is prescribed. For the specific device considered in this example, the Stokes drift decays significantly away from the oscillating walls such that the Eulerian and Lagrangian velocity are virtually indistinguishable from each other in the bulk of the microchannel. As such, the linear solver for the second-order system converges for prescription of various boundary conditions ($\vec{v}_2=\vec{0}$, $\vec{v}^\textrm{L}=\vec{0}$, or $\vec{v}^\textrm{M}=\vec{0}$) and yields almost identical results in the bulk of the microchannel. However, this is not the case for general acoustofluidic devices and different boundary conditions can lead to dissimilar results, as illustrated in the next section.

\subsubsection{Elliptical actuation}
\label{subsec: elliptical}
To illustrate the variations in the numerical solution arising from differences in boundary conditions, we now consider a modification of the example considered in Fig.~\ref{fig:rectilinear}. Specifically, we modify the boundary oscillation profile such that the bottom wall is prescribed a displacement of the form $\presdisp = (\d_{0x},\d_{0y})$ with
\begin{subequations} \label{eq:elliptical-actuation}
\begin{alignat}{2}
&\d_{0x}(t, x) =  0.6 u_0 \left[\sin \left(\frac{-2 \pi(x-W / 2)}{\lambda_s}+\omega t-\Delta \phi\right)
+\sin \left(\frac{-2 \pi(W / 2-x)}{\lambda_s}+\omega t\right)\right],
\\
& \d_{0y}(t, x) =  -u_0 \left[\cos \left(\frac{-2 \pi(x-W / 2)}{\lambda_s}+\omega t-\Delta \phi \right)
+\cos \left(\frac{-2 \pi(W / 2-x)}{\lambda_s}+\omega t\right)\right],
\end{alignat}
\end{subequations}
in which  $u_0 = 1.3\,\si{nm}$ is the displacement amplitude, $\omega = 2\pi f$ is the angular frequency with actuation frequency $f=1.97\,\si{MHz}$, $\lambda_s = W $ is the wavelength with $W=\SI{380}{\micro\metre}$ being the width of the microchannel, and $\Delta\phi = \pi/2$ is the phase angle. All other boundaries are assumed to be fixed. This choice of wall oscillation profile is motivated by surface acoustic wave devices~\cite{nama2015numerical,barnkob2018acoustically}. The channel dimensions and fluid parameters for this example are taken to be same as those considered in Fig.~\ref{fig:rectilinear}. A distinctive feature of the oscillation profile in Eq.~\eqref{eq:elliptical-actuation} is that it prescribes non-zero oscillations of the boundary in both $x$ and $y$ directions, in contrast to the rectilinear oscillations prescribed by Eq.~\eqref{eq:BAWoscillation}. For this profile, a point at the bottom boundary of the microchannel exhibits elliptical motion. This, in turn, results in a Stokes drift with non-zero components both along and perpendicular to the oscillating wall. 

To investigate the impact of second-order boundary conditions at the oscillating wall, we compare our numerical results for two different boundary conditions: (i) $\vec{v}^\textrm{L}=\vec{0}$ (Case A); and (ii) $\vec{v}^\textrm{M}=\vec{0}$ (Case B). To ensure the numerical consistency of these boundary conditions with the second-order mass balance Eq.~\eqref{eq:second-mass}, we handle the mass source term differently in each case. For the cases with boundary conditions $\vec{v}^\textrm{L}=\vec{0}$ and $\vec{v}^\textrm{M}=\vec{0}$, the second-order mass balance Eq.~\eqref{eq:second-mass} is reformulated as $\bnabla \cdot (\rho_0 \v^\textrm{L})= 0$ and $\bnabla \cdot (\rho_0 \v^\textrm{M}) = 0$, respectively.
These cases correspond to modeling choices in several previous modeling studies that employ $\vec{v}^\textrm{L}=\vec{0}$~\cite{kshetri2023acoustophoresis,kshetri2024evaluating,baasch2019acoustic,pavlic2022influence} or $\vec{v}^\textrm{M}=\vec{0}$~\cite{das2022extra,muller2014numerical,muller2015theoretical} as the boundary condition.
\begin{figure}
    \includegraphics[width=0.85\textwidth]{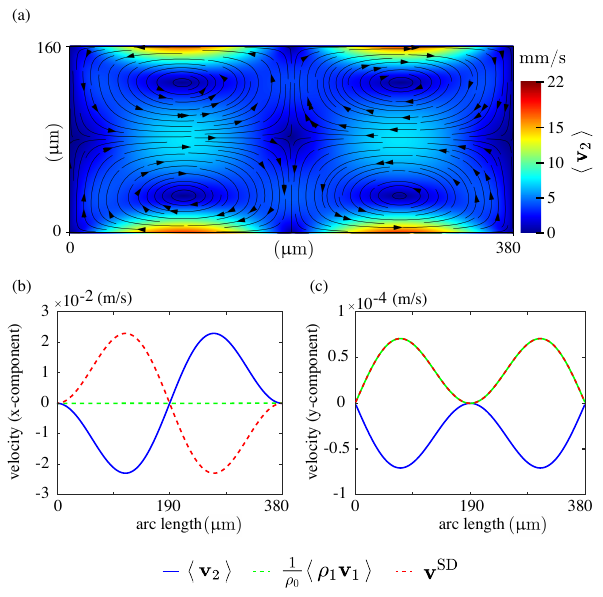}
    \centering
    \caption{(a) Streamlines and magnitude of the Eulerian streaming velocity $\langle\v_2\rangle$ solution obtained for the boundary oscillations corresponding to Eq.~\eqref{eq:elliptical-actuation} and $\v^\textrm{L}=\vec{0}$ boundary condition for the second-order problem (Case A). (b) and (c) plot the $x$ and $y$ component, respectively, of velocities $\v_2$, $\frac{1}{\rho_0 }\langle\rho_1 \v_1\rangle$, and $\v^{\textrm{SD}}$ at the bottom wall.}
    \label{fig:elliptical-with-SD-VL}
\end{figure}

\begin{figure}
    \includegraphics[width=0.85\textwidth]{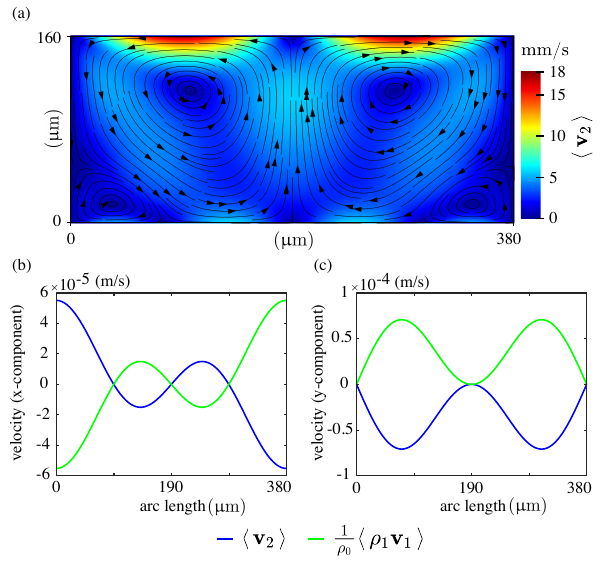}
    \centering
    \caption{(a) Streamlines and magnitude of the Eulerian streaming velocity $\langle\v_2\rangle$ solution obtained for the boundary oscillations corresponding to Eq.~\eqref{eq:elliptical-actuation} and $\v^\textrm{M}=\vec{0}$ boundary condition for the second-order problem (Case B). (b) and (c) plot the $x$ and $y$ component, respectively, of velocities $\v_2$, $\frac{1}{\rho_0 }\langle\rho_1 \v_1\rangle$, and $\v^{\textrm{SD}}$ at the bottom wall.}
    \label{fig:elliptical-no-SD-VM}
\end{figure}
Figs.~\ref{fig:elliptical-with-SD-VL} and~\ref{fig:elliptical-no-SD-VM} plot the numerical solutions corresponding to Cases A and B, respectively. Referring to Fig.~\ref{fig:elliptical-with-SD-VL}(a), the velocity field in Case A is characterized by four equal vortices, one in each quadrant of the microchannel. These vortices are qualitatively similar to those observed in Sec.~\ref{sec_rectilinear}. This is not surprising since the actuation frequency employed here corresponds to the half-wave resonance of the water-filled microchannel; as such, the impact of the choice of actuation profile on qualitative results is minimal. This observation is in excellent agreement with similar observations from Muller~\etal~\cite{muller2012numerical} for different actuation profiles. Moving on to Case B in Fig.~\ref{fig:elliptical-no-SD-VM}(a), significant differences are observed in the numerical solution compared to Case A. Specifically, the velocity field in Case B is characterized by two large vortices that span the entire height of the microchannel. In addition, there are two smaller vortices at the bottom corners of the microchannel. 

To further investigate the different representations of mean velocity at the oscillating boundary, Figs.~\ref{fig:elliptical-with-SD-VL}(b,c) plot the $x$ and $y$ components, respectively, of $\v_2$, $\v^\textrm{SD}$, and $\frac{1}{\rho_0}\langle \rho_1 \v_1 \rangle$ along the oscillating boundary (bottom wall) for Case A. Figs.~\ref{fig:elliptical-no-SD-VM}(b,c) plot the $x$ and $y$ components, respectively, of $\v_2$ and $\frac{1}{\rho_0}\langle \rho_1 \v_1 \rangle$ along the oscillating boundary (bottom wall) for Case B. Prior to the discussion of the results, we note that while the calculation of $\v_2$ requires the solution of the second-order system of equations, the quantities $\v^\textrm{SD}$ and $\frac{1}{\rho_0}\langle \rho_1 \v_1 \rangle$ are obtained directly through the first-order solution. Therefore, their value at the oscillating boundary can be calculated analytically from knowledge of the actuation profile in Eq.~\eqref{eq:elliptical-actuation}. The cases considered here differ only in their treatment of the second-order mass balance equation and boundary conditions; as such, they exhibit identical values of $\v^\textrm{SD}$ and $\frac{1}{\rho_0}\langle \rho_1 \vec{v}_1 \rangle$ at the oscillating boundary while the value of $\v_2$ is dictated by the prescribed boundary condition. Referring to Figs.~\ref{fig:elliptical-with-SD-VL}(b,c), it can be observed that for Case A, the value of $\v_2$ at the boundary is equal in magnitude and opposite in direction to $\v^\textrm{SD}$ to ensure a zero Lagrangian velocity $\v^\textrm{L}$ at the boundary. In contrast, for Case B (Figs.~\ref{fig:elliptical-no-SD-VM}(b,c)), $\v_2$ at the boundary is equal in magnitude and opposite in direction to $\frac{1}{\rho_0}\langle \rho_1 \v_1 \rangle$ to ensure a zero value of $\v^\textrm{M}$ at the boundary. The difference in the value of $\frac{1}{\rho_0}\langle \rho_1 \v_1 \rangle$ and $\v^\textrm{SD}$ explains the difference in results between Case A and Case B. Specifically, the different value $\frac{1}{\rho_0}\langle \rho_1 \v_1 \rangle$ and $\v^\textrm{SD}$ at the boundary implies that a different slip velocity is assigned to the Eulerian velocity in the two cases at the oscillating boundary which, in turn, impacts the results within the bulk of the microchannel. 

For the actuation profile in Eq.~\eqref{eq:elliptical-actuation}, the peak value of the $x$ component of $\v^\textrm{SD}$ is significantly larger than that of $\frac{1}{\rho_0}\langle \rho_1 \vec{v}_1 \rangle$ (see axes values in Figs.~\ref{fig:elliptical-with-SD-VL}(b,c)). Since these values are prescribed as slip velocities in Case A and B, respectively, the value of $\v_2$ at the oscillating boundary in Case A is significantly larger than that in Case B. Despite this, Eulerian streaming's global maximum value in both cases A and B is of the same order of magnitude. Lastly, comparing the $x$ and $y$ component plots in Figs.~\ref{fig:elliptical-with-SD-VL}(b,c), it can also be observed that the values of the $y$ component of $\v^\textrm{SD}$ is over two orders of magnitude smaller than the value of the $x$ component. In contrast, the values of $x$ and $y$ components of $\frac{1}{\rho_0}\langle \rho_1 \vec{v}_1 \rangle$ in Figs.~\ref{fig:elliptical-no-SD-VM}(b,c) are of similar magnitude. In Case B, the $y$ components of $\v^\textrm{SD}$ and $\v_2$ do not cancel each other, resulting in a non-zero $y$ component of $\v^\textrm{L}$ at the oscillating boundary and a non-physical loss of mass through the oscillating boundary. 
%
%

Having discussed the numerical results, we turn our attention to the physical interpretation of the boundary conditions in Cases A and B. As revealed by Eq.~\eqref{eq: vLvMrel_reform}, $\v^\textrm{M}$ and $\v^\textrm{L}$ differ by the term $\frac{1}{2 \rho_0 }\langle\ \bnabla \times(\rho_0 \bm{\xi}_1 \times \vone)\rangle$ which can, in general, be non-zero. At the oscillating boundary, this term is directly prescribed by the boundary oscillations and can be zero for some specific oscillation profiles. For instance, for rectilinear oscillation profiles, $\vec{\xi}_1$ and $\vec{v}_1$ are parallel and $\vec{\xi}_1 \times \vec{v}_1=0.$ Similarly, for oscillation profiles where $\bm{\xi}_1 \times \vone$ is non-zero, but uniform, the curl of this quantity $\bnabla \times(\bm{\xi}_1 \times \vone)$ is zero. However, for particles undergoing non-uniform elliptical oscillations as given in Eq.~\eqref{eq:elliptical-actuation}, the term $\langle\ \bnabla \times(\rho_0 \bm{\xi}_1 \times \vone)\rangle$ does not vanish, resulting in a non-zero difference between $\v^\textrm{M}$ and $\v^\textrm{L}$.

From a physical perspective, the no-slip boundary condition to be enforced at the oscillating surface is given by Eq.~\eqref{eq:noslip}. The boundary condition in Case A ($\v^\textrm{L}=0$) is obtained by a direct Lagrangian-to-Eulerian transformation through the expansion of Eq.~\eqref{eq:noslip} around the mean position of the oscillating boundary, and therefore, enforces no-slip at the oscillating boundary, correct up to $\mathcal{O}(\epsilon^2)$~\cite{bradley1996acoustic}. In contrast, the boundary condition in Case B ($\v^\textrm{M}=0$) is obtained by assuming that there is no Eulerian mass flux across the oscillating boundary. As discussed previously by Bradley~\cite{bradley1996acoustic}, this is an erroneous assumption and there can be non-zero mass transport across the oscillating surface in an Eulerian sense, which is sometimes referred to as ``McIntyre sink"~\cite{lighthill1978acoustic}. Depending on the prescribed oscillation profile, the difference between $\vec{v}^\textrm{M}$ and $\vec{v}^\textrm{L}$ may or may not yield significant differences in the numerical solution between Cases A and B, especially in the bulk of the microfluidic channel. Nonetheless, from a numerical perspective, as noted in Sec.~\ref{sec_matrix}, a non-zero value of $\vec{v}^\textrm{M}$ at the boundary may not be generally consistent with the second-order mass balance equation (Eq.~\eqref{eq:vLvM_div}) and may lead to non-convergence of the numerical solver. 

The results presented in this section reiterate the need to exercise prudence in enforcement of boundary conditions. Given that even boundary conditions that are inconsistent with governing equations can yield converged solutions---combined with the increasing use of new substrate materials in acoustofluidic devices that can have varying oscillation profiles---these results underscore the need for formal verification of numerical codes to identify such inconsistencies and avoid erroneous predictions.

\subsection{Spatially varying density field}
Next, we investigate the effect of considering fluid density as a spatially varying quantity. The spatial variation of fluid density might arise out of separate physical phenomena (e.g., due to concentration gradients~\cite{deshmukh2014acoustic,karlsen2018acoustic}, thermal gradients~\cite{chini2014large}, etc.). In this work, we do not examine the underlying physical phenomena that causes density variations; rather, density is prescribed as an analytical function of spatial coordinates. Specifically, we consider a linear variation of density along the $y$ direction as $\rho_0 = A + By$, in which $A$ and $B$ are constants. To facilitate comparisons against the constant density scenarios described in Sec.~\ref{sec_rectilinear}, $A$ is taken as the density of water, whereas $B$ is varied parametrically to investigate the impact of density gradient on the velocity profile. The prescribed oscillation profile as well as all other parameters are considered same as in Sec.~\ref{sec_rectilinear}.
\begin{figure}
    \centering
    \includegraphics[width=0.8\linewidth]{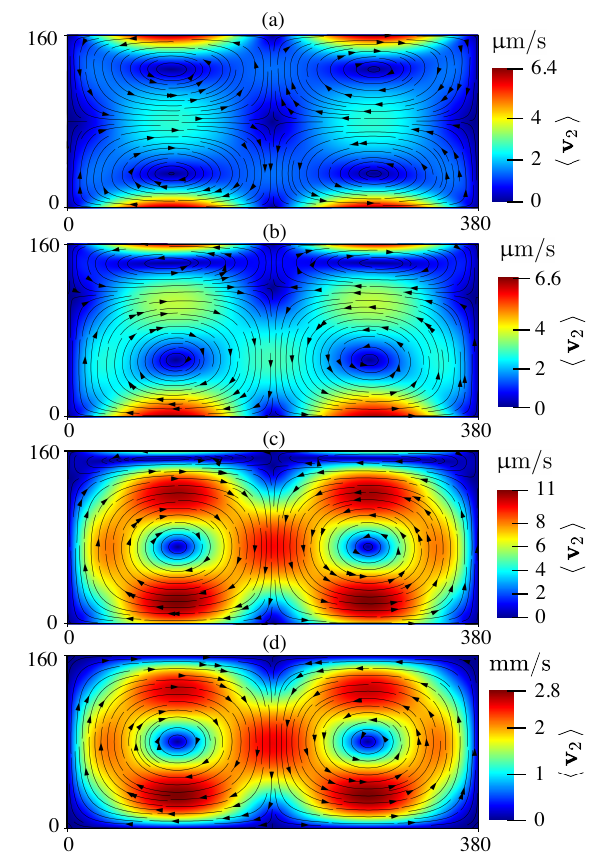}
    \caption{Streaming velocity field with variable density formulation. Density is prescribed as a spatial function of the form $\rho_0 = A + By$ with $A = 998$ ($\text{kg/m}^3$). (a-d) plots the Eulerian velocity streamlines obtained for different $B$ ($\text{kg/m}^4$) with values $0, 1\times 10^3, 4\times 10^3$ and $1\times 10^6$ respectively. The channel dimensions are in \SI{}{~\micro\metre}, while the color legends indicate the velocity magnitude.}
    \label{fig:variable-density-streaming}
\end{figure}

Fig.~\ref{fig:variable-density-streaming} plots the magnitude and streamlines of the Eulerian streaming velocity field for four different values of $B$. Fig.~\ref{fig:variable-density-streaming}(a) corresponds to  $B=0$, which is identical to the constant density case in Fig.~\ref{fig:rectilinear} and is characterized by four equal vortices, one in each quadrant of the microchannel. Each of these four vortices is driven by four counter-rotating vortices in the boundary layer, which are not directly visible in Fig.~\ref{fig:variable-density-streaming}(a) since the boundary layer thickness ($\delta_v$) for this actuation frequency is much smaller than the channel height ($\delta_v=0.38\,\si{\micro \metre}$). As the value of $B$ is increased in Figs.~\ref{fig:variable-density-streaming}(b,c) to introduce a density gradient along the $y$ direction, the velocity field becomes asymmetric around the horizontal mid-axis of the microchannel. The span of the bottom bulk vortices grows while the upper bulk vortices shrink. Moreover, the span of the boundary vortices both at the top and bottom wall becomes smaller than the constant density case. Further, velocity maxima in the lower half of the channel move away from the bottom boundary. This trend continues with a progressive increase in the value of $B$ until the upper vortices and boundary vortices disappear completely and the velocity field is characterized by only two vortices that span the entire height of the channel; see Fig.~\ref{fig:variable-density-streaming}(d). Further, the Eulerian velocity maxima now appear within the bulk of the microchannel instead of near the top and bottom walls. It is interesting to note that the maximum streaming velocity exhibits a monotonic trend with increasing density gradient and increases by approximately two order of magnitudes as B is changed from $B=4\times10^3$ to $B=1\times10^6$. While this represents a three order of magnitude increase in the density gradient, the change in maximum density is modest from $\rho_0^\textrm{max}=998.64\,\si{kg \cdot m^{-3}}$ (for $B=4\times10^3$) to $\rho_0^\textrm{max}=1158\,\si{kg \cdot m^{-3}}$ (for $B=1\times10^6$). \hglt{These results suggest that a significant increase in streaming velocity can be achieved through a density gradient within the channel. In this case, the presented density gradient should be regarded as an illustrative example for the numerical test case, representing an inhomogeneous fluid density that could, in principle, result from a solute concentration gradient~\cite{karlsen2016acoustic,deshmukh2014acoustic}.
}

\section{Conclusions and discussion}

A perturbation-based, variable-coefficient acoustofluidic solver with a second-order accurate finite difference/volume implementation is presented in this work. Starting with the compressible Navier-Stokes equations, we employ a perturbation approach to split the problem into a harmonic first-order and a time-averaged second-order system. We introduce fluid Lagrangian and mass transport velocities to illustrate two analytically equivalent but numerically distinct forms of second-order mass source terms. Further, we systematically relate these two velocities to clarify their distinction and to highlight that a zero Lagrangian velocity boundary condition is not, in general, equivalent to a zero mass transport velocity condition. 

Our numerical approach utilizes a finite difference/volume formulation with strong form implementation of both the governing equations and boundary conditions. To ensure numerical accuracy and assess convergence rates, we report formal verification of our implementation via manufactured solutions. In a series of test cases, we demonstrate that a zero mass transport velocity boundary condition can produce erroneous results and differ from those obtained with a zero Lagrangian velocity boundary condition. In contrast to the existing benchmark case by Muller~\etal~\cite{muller2012numerical,muller2013ultrasound}, which is insensitive to this distinction, the test cases presented in this work are sensitive to the choice of boundary conditions and therefore serve as better benchmarks.
\hglt{We note that a comparison of our results against a direct numerical simulation of Navier-Stokes equations would provide clear and definitive verification of the boundary conditions used in a perturbation-based solver. However, performing such simulations would require a time-dependent compressible flow solver with extremely small time steps to resolve high-frequency acoustic systems' time scales. A direct solution to the Navier-Stokes equations in an Eulerian framework (e.g., finite difference/volume) would require the prescription of boundary conditions at the undeformed location of the oscillating boundary. As such, the boundary condition in Eq.~\eqref{eq:noslip} should be expanded around the mean position to obtain a Robin type boundary condition at the (stationary) mesh boundary, as described in Eq.~\eqref{eq: TS_expansion}. More specifically, the boundary condition for the (unsplit) compressible flow solver would read as 
$\v (\r_0,t)   = \dot{\presdisp}(\r_0,t) - \bnabla \v (\r,t)\Bigr\rvert_{\r=\r_0}  \presdisp (\r_0,t) + \mathcal{O} (\epsilon^3).$}

This work presents a formally-verified, variable-coefficient acoustofluidic solver that paves the way for further numerical developments. Specifically, this solver represents the first essential step in our ongoing efforts towards developing a multiphase/multicomponent acoustofluidic solver capable of handling immersed bodies (cells, bubbles, particles) of distinct density and viscosity within acoustically actuated fluid-filled microfluidic devices. Future work will focus on incorporating complex geometries and bodies immersed in actuated fluid through immersed boundary methods~\cite{dillinger2024steerable,ren20193d,durrer2022robot,ahmed2016rotational,guo2015controlling,li2015standing,ozcelik2016acoustofluidic}. Due to the first-order solver, our acoustofluidic solver is currently limited in terms of scalability. Specifically, the use of a sparse direct solver for the coupled Helmholtz system is not a scalable approach to solving the first-order system. In contrast, our iterative solver for the second-order low Mach system is highly scalable thanks to the projection preconditioner; scalability results for our low Mach iterative solvers have been reported in our prior works~\cite{thirumalaisamy2023pre,thirumalaisamy2023lowmach,nangia2019robust,thirumalaisamy2025consistent}. 
We are not aware of a scalable solver for the first-order system. As such, there is an ongoing need for the computational acoustofluidic community to address this difficult linear algebra problem.

\section*{Acknowledgements}
This work was supported, in part, by the National Science Foundation (OAC-1931368, CBET-2234387, OIA-2229636, CBET-2407937, CBET-2407938) and the American Heart Association (23CDA1048343).
\appendix
\section{Proof of $\langle a(t)b(t) \rangle = \frac{1}{2} 
\operatorname{Re}\left(\widehat{a} \widehat{b}^*\right)$} \label{sec_<ab>}

Let $a$ and $b$ be two oscillating quantities of the form
\begin{subequations}
\begin{alignat}{2}
    a(t)&=\ahat e^{i \omega t},\\
    b(t)&=\bhat e^{i \omega t},
\end{alignat}
\end{subequations}
and having a time period $T$ and angular frequency $\omega = 2\pi/T$. Here, $\widehat{a} = a^\textrm{r} +\iota a^\textrm{i}$ and $\widehat{b} = b^\textrm{r} +\iota b^\textrm{i}$ denote the complex amplitudes of $a$ and $b$, respectively, and $e^{\iota \omega t} = \cos(\omega t) + \iota \sin(\omega t)$ denotes the complex exponential function, with the understanding that $\operatorname{Re}(a(t))$ and $\operatorname{Re}(b(t))$ represent the physically-relevant quantities of interest. The time-average of product of two oscillating quantities is expressed as

\begin{align}
\left\langle a(t)b(t)\right\rangle 
& := \frac{1}{T} \int_0^T \operatorname{Re}\left(\widehat{a} e^{\iota \omega t}\right) \operatorname{Re}\left(\widehat{b} e^{\iota \omega t}\right) \; \d t \nonumber \\
& = \frac{1}{T} \int_0^T \left(a^\textrm{r} \cos (\omega t)-a^\textrm{i} \sin (\omega t)\right)\left(b^\textrm{r} \cos (\omega t)-b^\textrm{i} \sin (\omega t)\right)\; \d t \nonumber \\
&= \frac{1}{T} \int_0^T \left(a^\textrm{r} b^\textrm{r} \cos ^2(\omega t)+a^\textrm{i} b^\textrm{i} \sin ^2(\omega t)\right) \; \d t
-\frac{1}{T} \int_0^T \left(a^\textrm{r} b^\textrm{i}+a^\textrm{i} b^\textrm{r}\right) \cos (\omega t) \sin (\omega t)\; \d t \nonumber \\
& = \frac{1}{2}\left(a^\textrm{r} b^\textrm{r} + a^\textrm{i} b^\textrm{i}\right) \nonumber \\
&= \frac{1}{2} \operatorname{Re}\left(\widehat{a} \widehat{b}^*\right). \label{eq:<ab>}
\end{align}
Here, $(\bullet)^*$ denotes the complex conjugate of the quantity $(\bullet)$.

\section{Proof of $\langle \partial_t (a \, b) \rangle = 0$}
\label{proof:timederivative_product}

Using Eq.~\eqref{eq:<ab>} we can express the time-average of product of two oscillating quantities $a$ and $\partial_t b$ in terms of their complex amplitudes as
\begin{alignat}{2}
   \langle a \,(\partial_t b) \rangle &= \frac{1}{2} \operatorname{Re} \left(\ahat (\iota\omega\, \bhat)^* \right) \nonumber \\
   &=\frac{1}{2} \textrm{Re}\left((a^{\rm r}+\iota a^{\rm i}) \{(i \omega )(b^{\rm r}+\iota b^{\rm i})\}^*\right) \nonumber \\
   &=\frac{1}{2} \textrm{Re}\left((a^{\rm r}+\iota a^{\rm i}) (-\iota \omega )(b^{\rm r} - \iota b^{\rm i})\right) \nonumber \\
   &=\frac{1}{2} \left(-a^{\rm r} b^{\rm i} + a^{\rm i} b^{\rm r}\right). \label{eq:a_dtb}
\end{alignat}
Similarly, it can be shown that 
\begin{equation}
    \langle (\partial_t a) \, b \rangle=\frac{1}{2} (-b^{\rm r} a^{\rm i} + b^{\rm i} a^{\rm r}).\label{eq:b_dta}
\end{equation} 
Adding Eqs.~\eqref{eq:a_dtb} and~\eqref{eq:b_dta} yields $\langle a \,(\partial_t b) \rangle + \langle (\partial_t a) \, b \rangle$ = 0, which implies that $\langle \partial_t (a \, b) \rangle = 0$.

\section{Separation of orders in the conservation of mass and momentum} \label{sec_sep_orders}
\subsection{Conservation of mass}
Substitution of the perturbation expansion in the conservation of mass Eq.~\eqref{eq:mass} yields
\begin{equation} 
    \frac{\partial }{\partial t} \bigg[ \rho_0 + \rho_1 + \rho_2 \bigg] + \bnabla \cdot \bigg [ (\rho_0 + \rho_1 + \rho_2)(\v_0+
    \v_1 + \v_2)\bigg] = 0,
\end{equation}
which can be re-arranged as
\begin{equation}
\underbrace{\frac{\partial \rho_0}{\partial t} + \bnabla \cdot (\rho_0 \v_0)
     }_{\textrm{Zeroth-order terms}} 
    + 
    \underbrace{
    \frac{\partial \rho_1}{\partial t}+ \bnabla \cdot [\rho_0 \v_1 + \rho_1 \v_0]
     }_{\textrm{First-order terms}} 
    + 
     \underbrace{
    \frac{\partial \rho_2}{\partial t}
    + \bnabla \cdot [\rho_0 \v_2 + \rho_2 \v_0 + \rho_1 \v_1]
     }_{\textrm{Second-order terms}} 
    = 0.
\end{equation}
Taking $\v_0$ to be zero (i.e., no flow in the absence of actuation), the conservation of mass equation at various orders of $\epsilon$ reads as
\begin{subequations}
\begin{alignat}{2}
&\textbf{Zeroth-order:} \qquad && \frac{\partial \rho_0}{\partial t}  = 0, \label{eq:zeroth-order-mass}\\
&\textbf{First-order:} \qquad &&  \frac{\partial \rho_1}{\partial t}+ \bnabla \cdot (\rho_0 \v_1)  = 0, \\
&\textbf{Second-order:} \qquad &&
    \frac{\partial \rho_2}{\partial t}+ \bnabla \cdot (\rho_0 \v_2 + \rho_1 \v_1) = 0.\label{eq:2ndcontinuity}
\end{alignat}    
\end{subequations}
Time-averaging the second-order mass balance equation gives
\begin{equation}
    \bnabla \cdot \langle\rho_0 \v_2\rangle=-\bnabla \cdot\langle\rho_1 \v_1\rangle.\label{eq:2ndcontinuity_2}
\end{equation}
\hglt{Note that the second-order quantities, in general, consist of a term proportional to $e^{2 i \omega t}$ and a time-independent component. Hence, the time average of the first term in Eq. \eqref{eq:2ndcontinuity} vanishes.}
\subsection{Conservation of linear momentum}
Substituting the perturbation expansion in Eq.~\eqref{eq:momentum}, we get
\begin{equation}
\begin{split}  
&\frac{\partial}{\partial t}[(\rho_0+\rho_1 +\rho_2)(\v_0+\v_1+\v_2)] \\
&+\bnabla \cdot[(\rho_0+\rho_1+\rho_2)\{(\v_0+\v_1+\v_2) \otimes(\v_0+\v_1+\v_2)\}]\\
 &= -\bnabla [p_0+p_1+p_2] 
+\bnabla \cdot [\mu \{ (\nabla \v_0+\nabla \v_1+\nabla \v_2)+(\nabla \v_0+\nabla \v_1+\nabla \v_2)^{\intercal}\}] \\
   & +\nabla[\lambda \bnabla \cdot(\v_0+\v_1+\v_2)].
\end{split}
\end{equation}
The above equation can be re-arranged as
\begin{equation}
\begin{split}
&\frac{\partial}{\partial t}[\rho_0 \v_0]+\bnabla \cdot[\rho_0 \v_0 \otimes \v_0] 
+\frac{\partial}{\partial t}[\rho_0 \v_1+\rho_1 \v_0]+\bnabla \cdot[\rho_0 \v_0 \otimes \v_1+\rho_0 \v_1 \otimes \v_0 +\rho_1 \v_0 \otimes \v_0] 
\\
&+\frac{\partial}{\partial t}[\rho_0 \v_2+\rho_2 \v_0+\rho_1 \v_1] 
+\bnabla \cdot[\rho_0(\v_0 \otimes \v_2+\v_2 \otimes \v_0+\v_1 \otimes \v_1)
\\
&+\rho_1(\v_0 \otimes \v_1+\v_1 \otimes \v_0)+\rho_2(\v_0 \otimes \v_0)] 
\\
&           = -\bnabla p_0+\bnabla \cdot[\mu \{\bnabla \v_0+(\bnabla \v_0)^\intercal\}]+\bnabla[\lambda \nabla \cdot \v_0] -\bnabla p_1+\\
&           \bnabla \cdot[\mu\{\bnabla \v_1+(\bnabla \v_1)^{\intercal}\}]+\bnabla[\lambda \bnabla \cdot \v_1] \\
&           \quad -\bnabla p_2+\bnabla \cdot[\mu\{\nabla \v_2+(\nabla \v_2)^{\intercal}\}]+\bnabla[\lambda \bnabla \cdot \v_2].
\end{split}
\end{equation}
Setting $\v_0$ to zero, the conservation of momentum equation at various orders of $\epsilon$ reads as
\begin{subequations}
\begin{alignat}{2}
&\textbf{Zeroth-order:} \qquad && \bnabla p_0 = 0, \\
&\textbf{First-order:} \qquad && \frac{\partial}{\partial t}(\rho_0 \v_1)=-\bnabla p_1+\bnabla \cdot[\mu\{\bnabla \v_1 +(\bnabla \v_1)^{\intercal}\}]+\bnabla[\lambda \bnabla \cdot \v_1], \\
&\textbf{Second-order:} \qquad && \frac{\partial}{\partial t}[\rho_0 \v_2 +\rho_1 \v_1]+\bnabla \cdot(\rho_0 \v_1 \otimes \v_1)= \nonumber\\
&\textbf{} \qquad &&-\bnabla p_2+\bnabla \cdot[\mu\{\bnabla \v_2+(\bnabla \v_2)^{\intercal}\}]+\bnabla[\lambda \bnabla \cdot \v_2].\label{eq:sec_momen}
\end{alignat}
\end{subequations}
Taking the time-average of second-order momentum equation gives
\begin{equation}
    \bnabla \cdot\langle\rho_0 \v_1 \otimes \v_1\rangle= -\bnabla \langle p_2 \rangle+\bnabla \cdot[\mu\{\bnabla \langle \v_2 \rangle+(\bnabla \langle \v_2\rangle)^{\intercal}\}]+\bnabla[\lambda \bnabla \cdot \langle\v_2\rangle].\label{eq:sec_momen2}
\end{equation}
In arriving at Eq.~\eqref{eq:sec_momen2} we have used the identity $\langle \frac{\partial}{\partial t}(\rho_1 \v_1) \rangle=0$; see Appendix~\ref{proof:timederivative_product}. \hglt{Further, $ \langle \frac{\partial}{\partial t}(\rho_0 \v_2) \rangle=0$ following the same argument as used in deriving Eq.~\eqref{eq:2ndcontinuity_2}}.
Lastly, since second-order quantities are assumed to be time-averaged, angle brackets around $\v_2$ and $p_2$ can be omitted in Eq.~\eqref{eq:sec_momen2}. 

\subsection{Equation of state} \label{sec:EoS}
The equation of state linking the fluid pressure to mass density can be expressed as
\begin{equation}
p=\pfunc(\rho).
\end{equation}
A Taylor series expansion can be used to express perturbations in fluid pressure caused by perturbations in density
\begin{equation}
\pfunc(\rho_0+\Delta \rho)=p(\rho_0) + \frac{\partial p}{\partial \rho}\bigg\lvert_{\rho=\rho_0} \Delta \rho + \frac{1}{2} \frac{\partial^2 p}{\partial \rho^2}\bigg\lvert_{\rho=\rho_0} (\Delta \rho)^2+....,
\end{equation}
in which $\Delta \rho=\rho_1 +\rho_2 + \mathcal{O}(\epsilon^3)$ represents the perturbations in the density around the equilibrium density $\rho_0$. By substituting the perturbation expansions of the pressure and density, we obtain
\begin{equation}
p_0 + p_1 + p_2 =p(\rho_0) + \frac{\partial p}{\partial \rho}\bigg\lvert_{\rho=\rho_0}  (\rho_1 +\rho_2 )+ \frac{1}{2} \frac{\partial^2 p}{\partial \rho^2}\bigg\lvert_{\rho=\rho_0} (\rho_1 +\rho_2)^2+\mathcal{O}(\epsilon^3),
\end{equation}
which can be rearranged as
\begin{equation}
p_0 + p_1 + p_2 =p(\rho_0) + \frac{\partial p}{\partial \rho}\bigg\lvert_{\rho=\rho_0}  \rho_1 + \frac{\partial p}{\partial \rho}\bigg\lvert_{\rho=\rho_0}  \rho_2+ \frac{1}{2} \frac{\partial^2 p}{\partial \rho^2}\bigg\lvert_{\rho=\rho_0} \rho_1^2+\mathcal{O}(\epsilon^3),
\end{equation}
in which we have retained terms upto second-order in $\epsilon$. Consequently, the equations of state at various orders of $\epsilon$ are as follows:
\begin{subequations}
\begin{alignat}{2}
& \textbf{Zeroth-order:} \qquad &&  p_0 = p(\rho_0), \\
& \textbf{First-order:} \qquad &&   p_1 = c_0^2  \rho_1, \qquad \textrm{with} \qquad c_0^2= \frac{\partial p}{\partial \rho}\bigg\lvert_{\rho=\rho_0}, \\
& \textbf{Second-order:} \qquad &&     p_2 = c_0^2  \rho_2 + \frac{1}{2} \frac{\partial^2 p}{\partial \rho^2}\bigg\lvert_{\rho=\rho_0} \rho_1^2.
\end{alignat}
\end{subequations}

\section{Proof of equivalence of Eqs.~\eqref{eq:vLvMrel} 
 and~\eqref{eq: vLvMrel_reform}}\label{proof:vl=vm+curl}
Consider the second term on the right hand side of Eq.~\eqref{eq: vLvMrel_reform} without the subscript `1' (for simplicity of notation)
\begin{subequations}
    \begin{alignat}{2}
    \frac{1}{2 \rho_0 }\langle &\bnabla \times(\rho_0 \bm{\xi} \times \v)\rangle =\frac{1}{2 \rho_0 }\langle 
    \varepsilon_{ijk} \varepsilon_{kpq} (\rho_0 \xi_p v_q)_{,j}
    \rangle \\
    &=\frac{1}{2 \rho_0 }\langle 
    (\delta_{ip}\delta_{jq}-\delta_{iq}\delta_{jp}) (\rho_0 \xi_p v_q)_{,j}
    \rangle\\
    &=\frac{1}{2 \rho_0 }\langle 
    (\xi_i \rho_0 v_j)_{,j}
    \rangle - \frac{1}{2 \rho_0 }\langle 
    (\rho_0 \xi_j  v_i)_{,j}
    \rangle
    \\
    &=\frac{1}{2 \rho_0 }\langle 
    \xi_i (\rho_0 v_j)_{,j}
    \rangle 
    + \frac{1}{2 \rho_0 }\langle 
    \xi_{i,j} \rho_0 v_j
    \rangle 
    -\frac{1}{2 \rho_0 }\langle 
    v_{i,j} \rho_0 \xi_j  
    \rangle
    -\frac{1}{2 \rho_0 }\langle 
    v_i (\rho_0 \xi_j)_{,j}  
    \rangle
    \\
    &=\frac{1}{2 \rho_0 }\langle 
    \xi_i (-\partial_t \rho_1)
    \rangle 
    + \frac{1}{2 \rho_0 }\langle 
    \xi_{i,j} \rho_0 \partial_t \xi_j
    \rangle 
    -\frac{1}{2 \rho_0 }\langle 
    v_{i,j} \rho_0 \xi_j  
    \rangle
    -\frac{1}{2 \rho_0 }\langle 
    \partial_t \xi_i (\rho_0 \xi_j)_{,j}  
    \rangle
    \\
    &=\frac{1}{2 \rho_0 }\langle 
    (\partial_t \xi_i) \rho_1
    \rangle 
     -\frac{1}{2 \rho_0 }\langle 
    \partial_t\xi_{i,j} \rho_0  \xi_j
    \rangle
    -\frac{1}{2 \rho_0 }\langle 
    v_{i,j} \rho_0 \xi_j  
    \rangle
    +\frac{1}{2 \rho_0 }\langle 
    \xi_i \partial_t (\rho_0 \xi_j)_{,j}  
    \rangle
    \\
    &=\frac{1}{2 \rho_0 }\langle 
    v_i \rho_1
    \rangle 
     -\frac{1}{2 }\langle 
    v_{i,j}  \xi_j
    \rangle
    -\frac{1}{2 }\langle 
    v_{i,j} \xi_j  
    \rangle
    +\frac{1}{2 \rho_0 }\langle 
    \xi_i (\rho_0 v_j)_{,j}  
    \rangle\\
    &=\frac{1}{2 \rho_0 }\langle 
    v_i \rho_1
    \rangle 
     -\langle 
    v_{i,j}  \xi_j
    \rangle
    +\frac{1}{2 \rho_0 }\langle 
    \xi_i (-\partial_t \rho_1)  
    \rangle\\
    &=\frac{1}{2 \rho_0 }\langle 
    v_i \rho_1
    \rangle 
     -\langle 
    v_{i,j}  \xi_j
    \rangle
    +\frac{1}{2 \rho_0 }\langle 
     (\partial_t \xi_i) \rho_1 
    \rangle \\
    &=\frac{1}{\rho_0 }\langle 
    v_i \rho_1
    \rangle 
     -\langle 
    v_{i,j}  \xi_j
    \rangle
    \end{alignat}
\end{subequations}
in which we have used $(\rho_0 v_j)_{,j} = -\partial_t \rho_1 $ from the first-order mass balance Eq.~\eqref{eq:first-mass} that has no mass source term, $\partial_t \rho_0 = 0$ from Eq.~\eqref{eq:zeroth-order-mass}, and $\langle a \,(\partial_t b) \rangle =- \langle (\partial_t a) \, b \rangle$ for two oscillating quantities $a$ and $b$ (see~\ref{proof:timederivative_product}). Therefore, 
\begin{equation}
    \frac{1}{2 \rho_0 }\langle \bnabla \times(\rho_0 \bm{\xi}_1 \times \vone)\rangle = \frac{1}{\rho_0}\langle \rhoone \vone \rangle - \langle(\bnabla{\vone})\bm{\xi}_1\rangle,
\end{equation}
which implies that Eqs.~\eqref{eq:vLvMrel} and~\eqref{eq: vLvMrel_reform} are equivalent.

\section{Convergence analysis of de-coupled second-order system}
\label{accuracy and convergence_decoupled}
\begin{figure}[h!]
    \centering
    \includegraphics[width=0.85\linewidth]{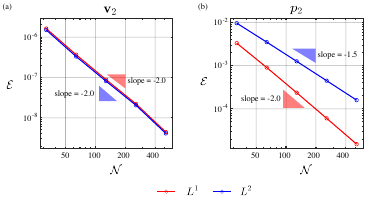}
    \caption{Spatial convergence rate of errors for the decoupled, second-order system with manufactured solutions. Each figure presents the log-log plot of $L^1$ (red curve) and $L^2$ (blue curve) error norms as a function of number of grid cells ($\mathcal{N}$) in each direction of the 2D $\mathcal{N}\times \mathcal{N}$ square domain. Figs. (a) and (b) plot the error norms for second-order velocity and pressure, respectively. The slope in each figure represents the spatial convergence rate of the error.}
    \label{fig:convergence_2nd_decoupled}
\end{figure}
To test the accuracy and spatial convergence rate of the decoupled second-order system, we follow the procedure of manufactured solutions described in Sec.~\ref{accuracy and convergence}. The first-order coupling terms are removed and Eq.~\eqref{eq:matrix-second-MS} is modified to
\begin{align} \label{eq:matrix-second-MS-decoupled}
\begin{bmatrix}
\V{L_{\mu+\lambda}} & \G\\
-\Drho_0 & \V{0}
\end{bmatrix}  
\begin{bmatrix}
\v_2 \\
\vptwo
\end{bmatrix} 
&=
\begin{bmatrix}
\vec{s}_{\v} \\
\vec{s_{p}} = \V{0}
\end{bmatrix}.
\end{align}
All relevant parameters are taken to be same as those considered in Sec.~\ref{accuracy and convergence}; see Eq.~\eqref{eq:fluid_parameters}. The manufactured solution for $\v_2$ and $p_2$ is chosen as 
\begin{subequations}
    \begin{alignat}{2}
        u_2 &= -x^3-y^3,\\ 
        v_2 &= -x^2-y^2,\\ 
        p_2 &= xy+x^2y^2.
    \end{alignat}
\end{subequations}
The manufactured second-order velocity and pressure fields are then substituted into Eq.~\eqref{eq:matrix-second-MS-decoupled} to determine the momentum source term $\vec{s}_{\v}$.
Figs.~\ref{fig:convergence_2nd_decoupled}(a,b) plot the $L^1$ and $L^2$ norm of errors in velocity and pressure as a function of $\mathcal{N}$, in which $\mathcal{N}$ represents the number of grid cells along each direction of the square domain. Second-order accuracy is observed for velocity in both the norms. For pressure, $L^1$ norm of error converges with second-order, whereas the $L^2$ norm converges at order 1.5. The reduction of accuracy for the $L^2$ norm of the pressure error is attributed to the spatially varying (shear) viscosity, which affects the interpolation accuracy at the domain corners. We recover second-order accuracy in the $L^2$ norm of the pressure error with spatially constant viscosity field; data not shown for brevity.   




\bibliography{bibliography.bib}

\providecommand{\noopsort}[1]{}\providecommand{\singleletter}[1]{#1}%
\begin{thebibliography}{61}%
\makeatletter
\providecommand \@ifxundefined [1]{%
 \@ifx{#1\undefined}
}%
\providecommand \@ifnum [1]{%
 \ifnum #1\expandafter \@firstoftwo
 \else \expandafter \@secondoftwo
 \fi
}%
\providecommand \@ifx [1]{%
 \ifx #1\expandafter \@firstoftwo
 \else \expandafter \@secondoftwo
 \fi
}%
\providecommand \natexlab [1]{#1}%
\providecommand \enquote  [1]{``#1''}%
\providecommand \bibnamefont  [1]{#1}%
\providecommand \bibfnamefont [1]{#1}%
\providecommand \citenamefont [1]{#1}%
\providecommand \href@noop [0]{\@secondoftwo}%
\providecommand \href [0]{\begingroup \@sanitize@url \@href}%
\providecommand \@href[1]{\@@startlink{#1}\@@href}%
\providecommand \@@href[1]{\endgroup#1\@@endlink}%
\providecommand \@sanitize@url [0]{\catcode `\\12\catcode `\$12\catcode `\&12\catcode `\#12\catcode `\^12\catcode `\_12\catcode `\%12\relax}%
\providecommand \@@startlink[1]{}%
\providecommand \@@endlink[0]{}%
\providecommand \url  [0]{\begingroup\@sanitize@url \@url }%
\providecommand \@url [1]{\endgroup\@href {#1}{\urlprefix }}%
\providecommand \urlprefix  [0]{URL }%
\providecommand \Eprint [0]{\href }%
\providecommand \doibase [0]{https://doi.org/}%
\providecommand \selectlanguage [0]{\@gobble}%
\providecommand \bibinfo  [0]{\@secondoftwo}%
\providecommand \bibfield  [0]{\@secondoftwo}%
\providecommand \translation [1]{[#1]}%
\providecommand \BibitemOpen [0]{}%
\providecommand \bibitemStop [0]{}%
\providecommand \bibitemNoStop [0]{.\EOS\space}%
\providecommand \EOS [0]{\spacefactor3000\relax}%
\providecommand \BibitemShut  [1]{\csname bibitem#1\endcsname}%
\let\auto@bib@innerbib\@empty
\bibitem [{\citenamefont {Laurell}\ and\ \citenamefont {Lenshof}(2014)}]{laurell2014microscale}%
  \BibitemOpen
  \bibfield  {author} {\bibinfo {author} {\bibfnamefont {T.}~\bibnamefont {Laurell}}\ and\ \bibinfo {author} {\bibfnamefont {A.}~\bibnamefont {Lenshof}},\ }\href@noop {} {\emph {\bibinfo {title} {Microscale acoustofluidics}}}\ (\bibinfo  {publisher} {Royal Society of Chemistry},\ \bibinfo {year} {2014})\BibitemShut {NoStop}%
\bibitem [{\citenamefont {Ding}\ \emph {et~al.}(2013)\citenamefont {Ding}, \citenamefont {Li}, \citenamefont {Lin}, \citenamefont {Stratton}, \citenamefont {Nama}, \citenamefont {Guo}, \citenamefont {Slotcavage}, \citenamefont {Mao}, \citenamefont {Shi}, \citenamefont {Costanzo} \emph {et~al.}}]{ding2013surface}%
  \BibitemOpen
  \bibfield  {author} {\bibinfo {author} {\bibfnamefont {X.}~\bibnamefont {Ding}}, \bibinfo {author} {\bibfnamefont {P.}~\bibnamefont {Li}}, \bibinfo {author} {\bibfnamefont {S.-C.~S.}\ \bibnamefont {Lin}}, \bibinfo {author} {\bibfnamefont {Z.~S.}\ \bibnamefont {Stratton}}, \bibinfo {author} {\bibfnamefont {N.}~\bibnamefont {Nama}}, \bibinfo {author} {\bibfnamefont {F.}~\bibnamefont {Guo}}, \bibinfo {author} {\bibfnamefont {D.}~\bibnamefont {Slotcavage}}, \bibinfo {author} {\bibfnamefont {X.}~\bibnamefont {Mao}}, \bibinfo {author} {\bibfnamefont {J.}~\bibnamefont {Shi}}, \bibinfo {author} {\bibfnamefont {F.}~\bibnamefont {Costanzo}}, \emph {et~al.},\ }\bibfield  {title} {\bibinfo {title} {Surface acoustic wave microfluidics},\ }\href@noop {} {\bibfield  {journal} {\bibinfo  {journal} {Lab on a Chip}\ }\textbf {\bibinfo {volume} {13}},\ \bibinfo {pages} {3626} (\bibinfo {year} {2013})}\BibitemShut {NoStop}%
\bibitem [{\citenamefont {Durrer}\ \emph {et~al.}(2022)\citenamefont {Durrer}, \citenamefont {Agrawal}, \citenamefont {Ozgul}, \citenamefont {Neuhauss}, \citenamefont {Nama},\ and\ \citenamefont {Ahmed}}]{durrer2022robot}%
  \BibitemOpen
  \bibfield  {author} {\bibinfo {author} {\bibfnamefont {J.}~\bibnamefont {Durrer}}, \bibinfo {author} {\bibfnamefont {P.}~\bibnamefont {Agrawal}}, \bibinfo {author} {\bibfnamefont {A.}~\bibnamefont {Ozgul}}, \bibinfo {author} {\bibfnamefont {S.~C.}\ \bibnamefont {Neuhauss}}, \bibinfo {author} {\bibfnamefont {N.}~\bibnamefont {Nama}},\ and\ \bibinfo {author} {\bibfnamefont {D.}~\bibnamefont {Ahmed}},\ }\bibfield  {title} {\bibinfo {title} {A robot-assisted acoustofluidic end effector},\ }\href@noop {} {\bibfield  {journal} {\bibinfo  {journal} {Nature Communications}\ }\textbf {\bibinfo {volume} {13}},\ \bibinfo {pages} {6370} (\bibinfo {year} {2022})}\BibitemShut {NoStop}%
\bibitem [{\citenamefont {Wei}\ \emph {et~al.}(2023)\citenamefont {Wei}, \citenamefont {Wang}, \citenamefont {Wang},\ and\ \citenamefont {Duan}}]{wei2023microscale}%
  \BibitemOpen
  \bibfield  {author} {\bibinfo {author} {\bibfnamefont {W.}~\bibnamefont {Wei}}, \bibinfo {author} {\bibfnamefont {Y.}~\bibnamefont {Wang}}, \bibinfo {author} {\bibfnamefont {Z.}~\bibnamefont {Wang}},\ and\ \bibinfo {author} {\bibfnamefont {X.}~\bibnamefont {Duan}},\ }\bibfield  {title} {\bibinfo {title} {Microscale acoustic streaming for biomedical and bioanalytical applications},\ }\href@noop {} {\bibfield  {journal} {\bibinfo  {journal} {TrAC Trends in Analytical Chemistry}\ ,\ \bibinfo {pages} {116958}} (\bibinfo {year} {2023})}\BibitemShut {NoStop}%
\bibitem [{\citenamefont {Huang}\ \emph {et~al.}(2015)\citenamefont {Huang}, \citenamefont {Ren}, \citenamefont {Nama}, \citenamefont {Li}, \citenamefont {Li}, \citenamefont {Yao}, \citenamefont {Cuento}, \citenamefont {Wei}, \citenamefont {Chen}, \citenamefont {Xie} \emph {et~al.}}]{huang2015acoustofluidic}%
  \BibitemOpen
  \bibfield  {author} {\bibinfo {author} {\bibfnamefont {P.-H.}\ \bibnamefont {Huang}}, \bibinfo {author} {\bibfnamefont {L.}~\bibnamefont {Ren}}, \bibinfo {author} {\bibfnamefont {N.}~\bibnamefont {Nama}}, \bibinfo {author} {\bibfnamefont {S.}~\bibnamefont {Li}}, \bibinfo {author} {\bibfnamefont {P.}~\bibnamefont {Li}}, \bibinfo {author} {\bibfnamefont {X.}~\bibnamefont {Yao}}, \bibinfo {author} {\bibfnamefont {R.~A.}\ \bibnamefont {Cuento}}, \bibinfo {author} {\bibfnamefont {C.-H.}\ \bibnamefont {Wei}}, \bibinfo {author} {\bibfnamefont {Y.}~\bibnamefont {Chen}}, \bibinfo {author} {\bibfnamefont {Y.}~\bibnamefont {Xie}}, \emph {et~al.},\ }\bibfield  {title} {\bibinfo {title} {An acoustofluidic sputum liquefier},\ }\href@noop {} {\bibfield  {journal} {\bibinfo  {journal} {Lab on a Chip}\ }\textbf {\bibinfo {volume} {15}},\ \bibinfo {pages} {3125} (\bibinfo {year} {2015})}\BibitemShut {NoStop}%
\bibitem [{\citenamefont {Ahmed}\ \emph {et~al.}(2013)\citenamefont {Ahmed}, \citenamefont {Chan}, \citenamefont {Lin}, \citenamefont {Muddana}, \citenamefont {Nama}, \citenamefont {Benkovic},\ and\ \citenamefont {Huang}}]{ahmed2013tunable}%
  \BibitemOpen
  \bibfield  {author} {\bibinfo {author} {\bibfnamefont {D.}~\bibnamefont {Ahmed}}, \bibinfo {author} {\bibfnamefont {C.~Y.}\ \bibnamefont {Chan}}, \bibinfo {author} {\bibfnamefont {S.-C.~S.}\ \bibnamefont {Lin}}, \bibinfo {author} {\bibfnamefont {H.~S.}\ \bibnamefont {Muddana}}, \bibinfo {author} {\bibfnamefont {N.}~\bibnamefont {Nama}}, \bibinfo {author} {\bibfnamefont {S.~J.}\ \bibnamefont {Benkovic}},\ and\ \bibinfo {author} {\bibfnamefont {T.~J.}\ \bibnamefont {Huang}},\ }\bibfield  {title} {\bibinfo {title} {Tunable, pulsatile chemical gradient generation via acoustically driven oscillating bubbles},\ }\href@noop {} {\bibfield  {journal} {\bibinfo  {journal} {Lab on a Chip}\ }\textbf {\bibinfo {volume} {13}},\ \bibinfo {pages} {328} (\bibinfo {year} {2013})}\BibitemShut {NoStop}%
\bibitem [{\citenamefont {Jun{\'a}Huang}\ \emph {et~al.}(2014)\citenamefont {Jun{\'a}Huang} \emph {et~al.}}]{junahuang2014reliable}%
  \BibitemOpen
  \bibfield  {author} {\bibinfo {author} {\bibfnamefont {T.}~\bibnamefont {Jun{\'a}Huang}} \emph {et~al.},\ }\bibfield  {title} {\bibinfo {title} {A reliable and programmable acoustofluidic pump powered by oscillating sharp-edge structures},\ }\href@noop {} {\bibfield  {journal} {\bibinfo  {journal} {Lab on a Chip}\ }\textbf {\bibinfo {volume} {14}},\ \bibinfo {pages} {4319} (\bibinfo {year} {2014})}\BibitemShut {NoStop}%
\bibitem [{\citenamefont {Huang}\ \emph {et~al.}(2013)\citenamefont {Huang}, \citenamefont {Xie}, \citenamefont {Ahmed}, \citenamefont {Rufo}, \citenamefont {Nama}, \citenamefont {Chen}, \citenamefont {Chan},\ and\ \citenamefont {Huang}}]{huang2013acoustofluidic}%
  \BibitemOpen
  \bibfield  {author} {\bibinfo {author} {\bibfnamefont {P.-H.}\ \bibnamefont {Huang}}, \bibinfo {author} {\bibfnamefont {Y.}~\bibnamefont {Xie}}, \bibinfo {author} {\bibfnamefont {D.}~\bibnamefont {Ahmed}}, \bibinfo {author} {\bibfnamefont {J.}~\bibnamefont {Rufo}}, \bibinfo {author} {\bibfnamefont {N.}~\bibnamefont {Nama}}, \bibinfo {author} {\bibfnamefont {Y.}~\bibnamefont {Chen}}, \bibinfo {author} {\bibfnamefont {C.~Y.}\ \bibnamefont {Chan}},\ and\ \bibinfo {author} {\bibfnamefont {T.~J.}\ \bibnamefont {Huang}},\ }\bibfield  {title} {\bibinfo {title} {An acoustofluidic micromixer based on oscillating sidewall sharp-edges},\ }\href@noop {} {\bibfield  {journal} {\bibinfo  {journal} {Lab on a Chip}\ }\textbf {\bibinfo {volume} {13}},\ \bibinfo {pages} {3847} (\bibinfo {year} {2013})}\BibitemShut {NoStop}%
\bibitem [{\citenamefont {Ozcelik}\ \emph {et~al.}(2014)\citenamefont {Ozcelik}, \citenamefont {Ahmed}, \citenamefont {Xie}, \citenamefont {Nama}, \citenamefont {Qu}, \citenamefont {Nawaz},\ and\ \citenamefont {Huang}}]{ozcelik2014acoustofluidic}%
  \BibitemOpen
  \bibfield  {author} {\bibinfo {author} {\bibfnamefont {A.}~\bibnamefont {Ozcelik}}, \bibinfo {author} {\bibfnamefont {D.}~\bibnamefont {Ahmed}}, \bibinfo {author} {\bibfnamefont {Y.}~\bibnamefont {Xie}}, \bibinfo {author} {\bibfnamefont {N.}~\bibnamefont {Nama}}, \bibinfo {author} {\bibfnamefont {Z.}~\bibnamefont {Qu}}, \bibinfo {author} {\bibfnamefont {A.~A.}\ \bibnamefont {Nawaz}},\ and\ \bibinfo {author} {\bibfnamefont {T.~J.}\ \bibnamefont {Huang}},\ }\bibfield  {title} {\bibinfo {title} {An acoustofluidic micromixer via bubble inception and cavitation from microchannel sidewalls},\ }\href@noop {} {\bibfield  {journal} {\bibinfo  {journal} {Analytical chemistry}\ }\textbf {\bibinfo {volume} {86}},\ \bibinfo {pages} {5083} (\bibinfo {year} {2014})}\BibitemShut {NoStop}%
\bibitem [{\citenamefont {Pavlic}\ \emph {et~al.}(2023)\citenamefont {Pavlic}, \citenamefont {Harshbarger}, \citenamefont {Rosenthaler}, \citenamefont {Snedeker},\ and\ \citenamefont {Dual}}]{pavlic2023sharp}%
  \BibitemOpen
  \bibfield  {author} {\bibinfo {author} {\bibfnamefont {A.}~\bibnamefont {Pavlic}}, \bibinfo {author} {\bibfnamefont {C.~L.}\ \bibnamefont {Harshbarger}}, \bibinfo {author} {\bibfnamefont {L.}~\bibnamefont {Rosenthaler}}, \bibinfo {author} {\bibfnamefont {J.~G.}\ \bibnamefont {Snedeker}},\ and\ \bibinfo {author} {\bibfnamefont {J.}~\bibnamefont {Dual}},\ }\bibfield  {title} {\bibinfo {title} {Sharp-edge-based acoustofluidic chip capable of programmable pumping, mixing, cell focusing, and trapping},\ }\href@noop {} {\bibfield  {journal} {\bibinfo  {journal} {Physics of Fluids}\ }\textbf {\bibinfo {volume} {35}} (\bibinfo {year} {2023})}\BibitemShut {NoStop}%
\bibitem [{\citenamefont {Beyer}(1999)}]{beyer1999sounds}%
  \BibitemOpen
  \bibfield  {author} {\bibinfo {author} {\bibfnamefont {R.~T.}\ \bibnamefont {Beyer}},\ }\href@noop {} {\emph {\bibinfo {title} {Sounds of our times: two hundred years of acoustics}}}\ (\bibinfo  {publisher} {Springer Science \& Business Media},\ \bibinfo {year} {1999})\BibitemShut {NoStop}%
\bibitem [{\citenamefont {Friend}\ and\ \citenamefont {Yeo}(2011)}]{friend2011microscale}%
  \BibitemOpen
  \bibfield  {author} {\bibinfo {author} {\bibfnamefont {J.}~\bibnamefont {Friend}}\ and\ \bibinfo {author} {\bibfnamefont {L.~Y.}\ \bibnamefont {Yeo}},\ }\bibfield  {title} {\bibinfo {title} {Microscale acoustofluidics: Microfluidics driven via acoustics and ultrasonics},\ }\href@noop {} {\bibfield  {journal} {\bibinfo  {journal} {Reviews of Modern Physics}\ }\textbf {\bibinfo {volume} {83}},\ \bibinfo {pages} {647} (\bibinfo {year} {2011})}\BibitemShut {NoStop}%
\bibitem [{\citenamefont {Nama}\ \emph {et~al.}(2014)\citenamefont {Nama}, \citenamefont {Huang}, \citenamefont {Huang},\ and\ \citenamefont {Costanzo}}]{nama2014investigation}%
  \BibitemOpen
  \bibfield  {author} {\bibinfo {author} {\bibfnamefont {N.}~\bibnamefont {Nama}}, \bibinfo {author} {\bibfnamefont {P.-H.}\ \bibnamefont {Huang}}, \bibinfo {author} {\bibfnamefont {T.~J.}\ \bibnamefont {Huang}},\ and\ \bibinfo {author} {\bibfnamefont {F.}~\bibnamefont {Costanzo}},\ }\bibfield  {title} {\bibinfo {title} {Investigation of acoustic streaming patterns around oscillating sharp edges},\ }\href@noop {} {\bibfield  {journal} {\bibinfo  {journal} {Lab on a Chip}\ }\textbf {\bibinfo {volume} {14}},\ \bibinfo {pages} {2824} (\bibinfo {year} {2014})}\BibitemShut {NoStop}%
\bibitem [{\citenamefont {Nama}\ \emph {et~al.}(2015)\citenamefont {Nama}, \citenamefont {Barnkob}, \citenamefont {Mao}, \citenamefont {K{\"a}hler}, \citenamefont {Costanzo},\ and\ \citenamefont {Huang}}]{nama2015numerical}%
  \BibitemOpen
  \bibfield  {author} {\bibinfo {author} {\bibfnamefont {N.}~\bibnamefont {Nama}}, \bibinfo {author} {\bibfnamefont {R.}~\bibnamefont {Barnkob}}, \bibinfo {author} {\bibfnamefont {Z.}~\bibnamefont {Mao}}, \bibinfo {author} {\bibfnamefont {C.~J.}\ \bibnamefont {K{\"a}hler}}, \bibinfo {author} {\bibfnamefont {F.}~\bibnamefont {Costanzo}},\ and\ \bibinfo {author} {\bibfnamefont {T.~J.}\ \bibnamefont {Huang}},\ }\bibfield  {title} {\bibinfo {title} {Numerical study of acoustophoretic motion of particles in a pdms microchannel driven by surface acoustic waves},\ }\href@noop {} {\bibfield  {journal} {\bibinfo  {journal} {Lab on a Chip}\ }\textbf {\bibinfo {volume} {15}},\ \bibinfo {pages} {2700} (\bibinfo {year} {2015})}\BibitemShut {NoStop}%
\bibitem [{\citenamefont {Muller}\ \emph {et~al.}(2013)\citenamefont {Muller}, \citenamefont {Rossi}, \citenamefont {Marin}, \citenamefont {Barnkob}, \citenamefont {Augustsson}, \citenamefont {Laurell}, \citenamefont {Kaehler},\ and\ \citenamefont {Bruus}}]{muller2013ultrasound}%
  \BibitemOpen
  \bibfield  {author} {\bibinfo {author} {\bibfnamefont {P.~B.}\ \bibnamefont {Muller}}, \bibinfo {author} {\bibfnamefont {M.}~\bibnamefont {Rossi}}, \bibinfo {author} {\bibfnamefont {A.}~\bibnamefont {Marin}}, \bibinfo {author} {\bibfnamefont {R.}~\bibnamefont {Barnkob}}, \bibinfo {author} {\bibfnamefont {P.}~\bibnamefont {Augustsson}}, \bibinfo {author} {\bibfnamefont {T.}~\bibnamefont {Laurell}}, \bibinfo {author} {\bibfnamefont {C.~J.}\ \bibnamefont {Kaehler}},\ and\ \bibinfo {author} {\bibfnamefont {H.}~\bibnamefont {Bruus}},\ }\bibfield  {title} {\bibinfo {title} {Ultrasound-induced acoustophoretic motion of microparticles in three dimensions},\ }\href@noop {} {\bibfield  {journal} {\bibinfo  {journal} {Physical Review E—Statistical, Nonlinear, and Soft Matter Physics}\ }\textbf {\bibinfo {volume} {88}},\ \bibinfo {pages} {023006} (\bibinfo {year} {2013})}\BibitemShut {NoStop}%
\bibitem [{\citenamefont {Devendran}\ \emph {et~al.}(2022)\citenamefont {Devendran}, \citenamefont {Collins},\ and\ \citenamefont {Neild}}]{devendran2022role}%
  \BibitemOpen
  \bibfield  {author} {\bibinfo {author} {\bibfnamefont {C.}~\bibnamefont {Devendran}}, \bibinfo {author} {\bibfnamefont {D.~J.}\ \bibnamefont {Collins}},\ and\ \bibinfo {author} {\bibfnamefont {A.}~\bibnamefont {Neild}},\ }\bibfield  {title} {\bibinfo {title} {The role of channel height and actuation method on particle manipulation in surface acoustic wave (saw)-driven microfluidic devices},\ }\href@noop {} {\bibfield  {journal} {\bibinfo  {journal} {Microfluidics and Nanofluidics}\ }\textbf {\bibinfo {volume} {26}},\ \bibinfo {pages} {9} (\bibinfo {year} {2022})}\BibitemShut {NoStop}%
\bibitem [{\citenamefont {Vanneste}\ and\ \citenamefont {B{\"u}hler}(2011)}]{vanneste2011streaming}%
  \BibitemOpen
  \bibfield  {author} {\bibinfo {author} {\bibfnamefont {J.}~\bibnamefont {Vanneste}}\ and\ \bibinfo {author} {\bibfnamefont {O.}~\bibnamefont {B{\"u}hler}},\ }\bibfield  {title} {\bibinfo {title} {Streaming by leaky surface acoustic waves},\ }\href@noop {} {\bibfield  {journal} {\bibinfo  {journal} {Proceedings of the Royal Society A: Mathematical, Physical and Engineering Sciences}\ }\textbf {\bibinfo {volume} {467}},\ \bibinfo {pages} {1779} (\bibinfo {year} {2011})}\BibitemShut {NoStop}%
\bibitem [{\citenamefont {Orosco}\ and\ \citenamefont {Friend}(2022)}]{orosco2022modeling}%
  \BibitemOpen
  \bibfield  {author} {\bibinfo {author} {\bibfnamefont {J.}~\bibnamefont {Orosco}}\ and\ \bibinfo {author} {\bibfnamefont {J.}~\bibnamefont {Friend}},\ }\bibfield  {title} {\bibinfo {title} {Modeling fast acoustic streaming: Steady-state and transient flow solutions},\ }\href@noop {} {\bibfield  {journal} {\bibinfo  {journal} {Physical Review E}\ }\textbf {\bibinfo {volume} {106}},\ \bibinfo {pages} {045101} (\bibinfo {year} {2022})}\BibitemShut {NoStop}%
\bibitem [{\citenamefont {Lei}\ \emph {et~al.}(2017)\citenamefont {Lei}, \citenamefont {Hill},\ and\ \citenamefont {Glynne-Jones}}]{lei2017transducer}%
  \BibitemOpen
  \bibfield  {author} {\bibinfo {author} {\bibfnamefont {J.}~\bibnamefont {Lei}}, \bibinfo {author} {\bibfnamefont {M.}~\bibnamefont {Hill}},\ and\ \bibinfo {author} {\bibfnamefont {P.}~\bibnamefont {Glynne-Jones}},\ }\bibfield  {title} {\bibinfo {title} {Transducer-plane streaming patterns in thin-layer acoustofluidic devices},\ }\href@noop {} {\bibfield  {journal} {\bibinfo  {journal} {Physical Review Applied}\ }\textbf {\bibinfo {volume} {8}},\ \bibinfo {pages} {014018} (\bibinfo {year} {2017})}\BibitemShut {NoStop}%
\bibitem [{\citenamefont {Baasch}\ and\ \citenamefont {Dual}(2018)}]{baasch2018acoustofluidic}%
  \BibitemOpen
  \bibfield  {author} {\bibinfo {author} {\bibfnamefont {T.}~\bibnamefont {Baasch}}\ and\ \bibinfo {author} {\bibfnamefont {J.}~\bibnamefont {Dual}},\ }\bibfield  {title} {\bibinfo {title} {Acoustofluidic particle dynamics: Beyond the rayleigh limit},\ }\href@noop {} {\bibfield  {journal} {\bibinfo  {journal} {The Journal of the Acoustical Society of America}\ }\textbf {\bibinfo {volume} {143}},\ \bibinfo {pages} {509} (\bibinfo {year} {2018})}\BibitemShut {NoStop}%
\bibitem [{\citenamefont {Sankaranarayanan}\ \emph {et~al.}(2008)\citenamefont {Sankaranarayanan}, \citenamefont {Cular}, \citenamefont {Bhethanabotla},\ and\ \citenamefont {Joseph}}]{sankaranarayanan2008flow}%
  \BibitemOpen
  \bibfield  {author} {\bibinfo {author} {\bibfnamefont {S.~K.}\ \bibnamefont {Sankaranarayanan}}, \bibinfo {author} {\bibfnamefont {S.}~\bibnamefont {Cular}}, \bibinfo {author} {\bibfnamefont {V.~R.}\ \bibnamefont {Bhethanabotla}},\ and\ \bibinfo {author} {\bibfnamefont {B.}~\bibnamefont {Joseph}},\ }\bibfield  {title} {\bibinfo {title} {Flow induced by acoustic streaming on surface-acoustic-wave devices and its application in biofouling removal: A computational study and comparisons to experiment},\ }\href@noop {} {\bibfield  {journal} {\bibinfo  {journal} {Physical Review E—Statistical, Nonlinear, and Soft Matter Physics}\ }\textbf {\bibinfo {volume} {77}},\ \bibinfo {pages} {066308} (\bibinfo {year} {2008})}\BibitemShut {NoStop}%
\bibitem [{\citenamefont {Devendran}\ \emph {et~al.}(2014)\citenamefont {Devendran}, \citenamefont {Gralinski},\ and\ \citenamefont {Neild}}]{devendran2014separation}%
  \BibitemOpen
  \bibfield  {author} {\bibinfo {author} {\bibfnamefont {C.}~\bibnamefont {Devendran}}, \bibinfo {author} {\bibfnamefont {I.}~\bibnamefont {Gralinski}},\ and\ \bibinfo {author} {\bibfnamefont {A.}~\bibnamefont {Neild}},\ }\bibfield  {title} {\bibinfo {title} {Separation of particles using acoustic streaming and radiation forces in an open microfluidic channel},\ }\href@noop {} {\bibfield  {journal} {\bibinfo  {journal} {Microfluidics and nanofluidics}\ }\textbf {\bibinfo {volume} {17}},\ \bibinfo {pages} {879} (\bibinfo {year} {2014})}\BibitemShut {NoStop}%
\bibitem [{\citenamefont {Baasch}\ \emph {et~al.}(2020)\citenamefont {Baasch}, \citenamefont {Doinikov},\ and\ \citenamefont {Dual}}]{baasch2020acoustic}%
  \BibitemOpen
  \bibfield  {author} {\bibinfo {author} {\bibfnamefont {T.}~\bibnamefont {Baasch}}, \bibinfo {author} {\bibfnamefont {A.~A.}\ \bibnamefont {Doinikov}},\ and\ \bibinfo {author} {\bibfnamefont {J.}~\bibnamefont {Dual}},\ }\bibfield  {title} {\bibinfo {title} {Acoustic streaming outside and inside a fluid particle undergoing monopole and dipole oscillations},\ }\href@noop {} {\bibfield  {journal} {\bibinfo  {journal} {Physical Review E}\ }\textbf {\bibinfo {volume} {101}},\ \bibinfo {pages} {013108} (\bibinfo {year} {2020})}\BibitemShut {NoStop}%
\bibitem [{\citenamefont {Muller}\ \emph {et~al.}(2012)\citenamefont {Muller}, \citenamefont {Barnkob}, \citenamefont {Jensen},\ and\ \citenamefont {Bruus}}]{muller2012numerical}%
  \BibitemOpen
  \bibfield  {author} {\bibinfo {author} {\bibfnamefont {P.~B.}\ \bibnamefont {Muller}}, \bibinfo {author} {\bibfnamefont {R.}~\bibnamefont {Barnkob}}, \bibinfo {author} {\bibfnamefont {M.~J.~H.}\ \bibnamefont {Jensen}},\ and\ \bibinfo {author} {\bibfnamefont {H.}~\bibnamefont {Bruus}},\ }\bibfield  {title} {\bibinfo {title} {A numerical study of microparticle acoustophoresis driven by acoustic radiation forces and streaming-induced drag forces},\ }\href@noop {} {\bibfield  {journal} {\bibinfo  {journal} {Lab on a Chip}\ }\textbf {\bibinfo {volume} {12}},\ \bibinfo {pages} {4617} (\bibinfo {year} {2012})}\BibitemShut {NoStop}%
\bibitem [{\citenamefont {Das}\ \emph {et~al.}(2019)\citenamefont {Das}, \citenamefont {Snider},\ and\ \citenamefont {Bhethanabotla}}]{das2019acoustothermal}%
  \BibitemOpen
  \bibfield  {author} {\bibinfo {author} {\bibfnamefont {P.~K.}\ \bibnamefont {Das}}, \bibinfo {author} {\bibfnamefont {A.~D.}\ \bibnamefont {Snider}},\ and\ \bibinfo {author} {\bibfnamefont {V.~R.}\ \bibnamefont {Bhethanabotla}},\ }\bibfield  {title} {\bibinfo {title} {Acoustothermal heating in surface acoustic wave driven microchannel flow},\ }\href@noop {} {\bibfield  {journal} {\bibinfo  {journal} {Physics of Fluids}\ }\textbf {\bibinfo {volume} {31}} (\bibinfo {year} {2019})}\BibitemShut {NoStop}%
\bibitem [{\citenamefont {Ghorbani~Kharaji}\ \emph {et~al.}(2022)\citenamefont {Ghorbani~Kharaji}, \citenamefont {Kalantar},\ and\ \citenamefont {Bayareh}}]{ghorbani2022acoustic}%
  \BibitemOpen
  \bibfield  {author} {\bibinfo {author} {\bibfnamefont {Z.}~\bibnamefont {Ghorbani~Kharaji}}, \bibinfo {author} {\bibfnamefont {V.}~\bibnamefont {Kalantar}},\ and\ \bibinfo {author} {\bibfnamefont {M.}~\bibnamefont {Bayareh}},\ }\bibfield  {title} {\bibinfo {title} {Acoustic sharp-edge-based micromixer: a numerical study},\ }\href@noop {} {\bibfield  {journal} {\bibinfo  {journal} {Chemical Papers}\ }\textbf {\bibinfo {volume} {76}},\ \bibinfo {pages} {1721} (\bibinfo {year} {2022})}\BibitemShut {NoStop}%
\bibitem [{\citenamefont {Das}\ and\ \citenamefont {Bhethanabotla}(2022)}]{das2022extra}%
  \BibitemOpen
  \bibfield  {author} {\bibinfo {author} {\bibfnamefont {P.~K.}\ \bibnamefont {Das}}\ and\ \bibinfo {author} {\bibfnamefont {V.~R.}\ \bibnamefont {Bhethanabotla}},\ }\bibfield  {title} {\bibinfo {title} {Extra stress-mediated acoustic streaming in a surface acoustic wave driven microchannel filled with second-order fluids},\ }\href@noop {} {\bibfield  {journal} {\bibinfo  {journal} {Physical Review Fluids}\ }\textbf {\bibinfo {volume} {7}},\ \bibinfo {pages} {074404} (\bibinfo {year} {2022})}\BibitemShut {NoStop}%
\bibitem [{\citenamefont {Muller}\ and\ \citenamefont {Bruus}(2014)}]{muller2014numerical}%
  \BibitemOpen
  \bibfield  {author} {\bibinfo {author} {\bibfnamefont {P.~B.}\ \bibnamefont {Muller}}\ and\ \bibinfo {author} {\bibfnamefont {H.}~\bibnamefont {Bruus}},\ }\bibfield  {title} {\bibinfo {title} {Numerical study of thermoviscous effects in ultrasound-induced acoustic streaming in microchannels},\ }\href@noop {} {\bibfield  {journal} {\bibinfo  {journal} {Physical Review E}\ }\textbf {\bibinfo {volume} {90}},\ \bibinfo {pages} {043016} (\bibinfo {year} {2014})}\BibitemShut {NoStop}%
\bibitem [{\citenamefont {Muller}\ and\ \citenamefont {Bruus}(2015)}]{muller2015theoretical}%
  \BibitemOpen
  \bibfield  {author} {\bibinfo {author} {\bibfnamefont {P.~B.}\ \bibnamefont {Muller}}\ and\ \bibinfo {author} {\bibfnamefont {H.}~\bibnamefont {Bruus}},\ }\bibfield  {title} {\bibinfo {title} {Theoretical study of time-dependent, ultrasound-induced acoustic streaming in microchannels},\ }\href@noop {} {\bibfield  {journal} {\bibinfo  {journal} {Physical Review E}\ }\textbf {\bibinfo {volume} {92}},\ \bibinfo {pages} {063018} (\bibinfo {year} {2015})}\BibitemShut {NoStop}%
\bibitem [{\citenamefont {Kshetri}\ and\ \citenamefont {Nama}(2023)}]{kshetri2023acoustophoresis}%
  \BibitemOpen
  \bibfield  {author} {\bibinfo {author} {\bibfnamefont {K.~G.}\ \bibnamefont {Kshetri}}\ and\ \bibinfo {author} {\bibfnamefont {N.}~\bibnamefont {Nama}},\ }\bibfield  {title} {\bibinfo {title} {Acoustophoresis around an elastic scatterer in a standing wave field},\ }\href@noop {} {\bibfield  {journal} {\bibinfo  {journal} {Physical Review E}\ }\textbf {\bibinfo {volume} {108}},\ \bibinfo {pages} {045102} (\bibinfo {year} {2023})}\BibitemShut {NoStop}%
\bibitem [{\citenamefont {Kshetri}\ and\ \citenamefont {Nama}(2024)}]{kshetri2024evaluating}%
  \BibitemOpen
  \bibfield  {author} {\bibinfo {author} {\bibfnamefont {K.~G.}\ \bibnamefont {Kshetri}}\ and\ \bibinfo {author} {\bibfnamefont {N.}~\bibnamefont {Nama}},\ }\bibfield  {title} {\bibinfo {title} {Evaluating impedance boundary conditions to model interfacial dynamics in acoustofluidics},\ }\href@noop {} {\bibfield  {journal} {\bibinfo  {journal} {Physics of Fluids}\ }\textbf {\bibinfo {volume} {36}} (\bibinfo {year} {2024})}\BibitemShut {NoStop}%
\bibitem [{\citenamefont {Baasch}\ \emph {et~al.}(2019)\citenamefont {Baasch}, \citenamefont {Pavlic},\ and\ \citenamefont {Dual}}]{baasch2019acoustic}%
  \BibitemOpen
  \bibfield  {author} {\bibinfo {author} {\bibfnamefont {T.}~\bibnamefont {Baasch}}, \bibinfo {author} {\bibfnamefont {A.}~\bibnamefont {Pavlic}},\ and\ \bibinfo {author} {\bibfnamefont {J.}~\bibnamefont {Dual}},\ }\bibfield  {title} {\bibinfo {title} {Acoustic radiation force acting on a heavy particle in a standing wave can be dominated by the acoustic microstreaming},\ }\href@noop {} {\bibfield  {journal} {\bibinfo  {journal} {Physical Review E}\ }\textbf {\bibinfo {volume} {100}},\ \bibinfo {pages} {061102} (\bibinfo {year} {2019})}\BibitemShut {NoStop}%
\bibitem [{\citenamefont {Pavlic}\ \emph {et~al.}(2022)\citenamefont {Pavlic}, \citenamefont {Nagpure}, \citenamefont {Ermanni},\ and\ \citenamefont {Dual}}]{pavlic2022influence}%
  \BibitemOpen
  \bibfield  {author} {\bibinfo {author} {\bibfnamefont {A.}~\bibnamefont {Pavlic}}, \bibinfo {author} {\bibfnamefont {P.}~\bibnamefont {Nagpure}}, \bibinfo {author} {\bibfnamefont {L.}~\bibnamefont {Ermanni}},\ and\ \bibinfo {author} {\bibfnamefont {J.}~\bibnamefont {Dual}},\ }\bibfield  {title} {\bibinfo {title} {Influence of particle shape and material on the acoustic radiation force and microstreaming in a standing wave},\ }\href@noop {} {\bibfield  {journal} {\bibinfo  {journal} {Physical Review E}\ }\textbf {\bibinfo {volume} {106}},\ \bibinfo {pages} {015105} (\bibinfo {year} {2022})}\BibitemShut {NoStop}%
\bibitem [{\citenamefont {Bradley}(1996)}]{bradley1996acoustic}%
  \BibitemOpen
  \bibfield  {author} {\bibinfo {author} {\bibfnamefont {C.}~\bibnamefont {Bradley}},\ }\bibfield  {title} {\bibinfo {title} {Acoustic streaming field structure: The influence of the radiator},\ }\href@noop {} {\bibfield  {journal} {\bibinfo  {journal} {The Journal of the Acoustical Society of America}\ }\textbf {\bibinfo {volume} {100}},\ \bibinfo {pages} {1399} (\bibinfo {year} {1996})}\BibitemShut {NoStop}%
\bibitem [{\citenamefont {Nama}\ \emph {et~al.}(2016)\citenamefont {Nama}, \citenamefont {Huang}, \citenamefont {Huang},\ and\ \citenamefont {Costanzo}}]{nama2016investigation}%
  \BibitemOpen
  \bibfield  {author} {\bibinfo {author} {\bibfnamefont {N.}~\bibnamefont {Nama}}, \bibinfo {author} {\bibfnamefont {P.-H.}\ \bibnamefont {Huang}}, \bibinfo {author} {\bibfnamefont {T.~J.}\ \bibnamefont {Huang}},\ and\ \bibinfo {author} {\bibfnamefont {F.}~\bibnamefont {Costanzo}},\ }\bibfield  {title} {\bibinfo {title} {Investigation of micromixing by acoustically oscillated sharp-edges},\ }\href@noop {} {\bibfield  {journal} {\bibinfo  {journal} {Biomicrofluidics}\ }\textbf {\bibinfo {volume} {10}} (\bibinfo {year} {2016})}\BibitemShut {NoStop}%
\bibitem [{\citenamefont {Deshmukh}\ \emph {et~al.}(2014)\citenamefont {Deshmukh}, \citenamefont {Brzozka}, \citenamefont {Laurell},\ and\ \citenamefont {Augustsson}}]{deshmukh2014acoustic}%
  \BibitemOpen
  \bibfield  {author} {\bibinfo {author} {\bibfnamefont {S.}~\bibnamefont {Deshmukh}}, \bibinfo {author} {\bibfnamefont {Z.}~\bibnamefont {Brzozka}}, \bibinfo {author} {\bibfnamefont {T.}~\bibnamefont {Laurell}},\ and\ \bibinfo {author} {\bibfnamefont {P.}~\bibnamefont {Augustsson}},\ }\bibfield  {title} {\bibinfo {title} {Acoustic radiation forces at liquid interfaces impact the performance of acoustophoresis},\ }\href@noop {} {\bibfield  {journal} {\bibinfo  {journal} {Lab on a Chip}\ }\textbf {\bibinfo {volume} {14}},\ \bibinfo {pages} {3394} (\bibinfo {year} {2014})}\BibitemShut {NoStop}%
\bibitem [{\citenamefont {Karlsen}\ \emph {et~al.}(2016)\citenamefont {Karlsen}, \citenamefont {Augustsson},\ and\ \citenamefont {Bruus}}]{karlsen2016acoustic}%
  \BibitemOpen
  \bibfield  {author} {\bibinfo {author} {\bibfnamefont {J.~T.}\ \bibnamefont {Karlsen}}, \bibinfo {author} {\bibfnamefont {P.}~\bibnamefont {Augustsson}},\ and\ \bibinfo {author} {\bibfnamefont {H.}~\bibnamefont {Bruus}},\ }\bibfield  {title} {\bibinfo {title} {Acoustic force density acting on inhomogeneous fluids in acoustic fields},\ }\href@noop {} {\bibfield  {journal} {\bibinfo  {journal} {Physical review letters}\ }\textbf {\bibinfo {volume} {117}},\ \bibinfo {pages} {114504} (\bibinfo {year} {2016})}\BibitemShut {NoStop}%
\bibitem [{\citenamefont {Karlsen}\ \emph {et~al.}(2018)\citenamefont {Karlsen}, \citenamefont {Qiu}, \citenamefont {Augustsson},\ and\ \citenamefont {Bruus}}]{karlsen2018acoustic}%
  \BibitemOpen
  \bibfield  {author} {\bibinfo {author} {\bibfnamefont {J.~T.}\ \bibnamefont {Karlsen}}, \bibinfo {author} {\bibfnamefont {W.}~\bibnamefont {Qiu}}, \bibinfo {author} {\bibfnamefont {P.}~\bibnamefont {Augustsson}},\ and\ \bibinfo {author} {\bibfnamefont {H.}~\bibnamefont {Bruus}},\ }\bibfield  {title} {\bibinfo {title} {Acoustic streaming and its suppression in inhomogeneous fluids},\ }\href@noop {} {\bibfield  {journal} {\bibinfo  {journal} {Physical review letters}\ }\textbf {\bibinfo {volume} {120}},\ \bibinfo {pages} {054501} (\bibinfo {year} {2018})}\BibitemShut {NoStop}%
\bibitem [{\citenamefont {Bruus}(2012)}]{bruus2012acoustofluidics}%
  \BibitemOpen
  \bibfield  {author} {\bibinfo {author} {\bibfnamefont {H.}~\bibnamefont {Bruus}},\ }\bibfield  {title} {\bibinfo {title} {Acoustofluidics 2: Perturbation theory and ultrasound resonance modes},\ }\href@noop {} {\bibfield  {journal} {\bibinfo  {journal} {Lab on a Chip}\ }\textbf {\bibinfo {volume} {12}},\ \bibinfo {pages} {20} (\bibinfo {year} {2012})}\BibitemShut {NoStop}%
\bibitem [{\citenamefont {B{\"u}hler}(2014)}]{buhler2014waves}%
  \BibitemOpen
  \bibfield  {author} {\bibinfo {author} {\bibfnamefont {O.}~\bibnamefont {B{\"u}hler}},\ }\href@noop {} {\emph {\bibinfo {title} {Waves and mean flows}}}\ (\bibinfo  {publisher} {Cambridge University Press},\ \bibinfo {year} {2014})\BibitemShut {NoStop}%
\bibitem [{\citenamefont {Amestoy}\ \emph {et~al.}(2001)\citenamefont {Amestoy}, \citenamefont {Duff}, \citenamefont {Koster},\ and\ \citenamefont {L'Excellent}}]{MUMPS:1}%
  \BibitemOpen
  \bibfield  {author} {\bibinfo {author} {\bibfnamefont {P.}~\bibnamefont {Amestoy}}, \bibinfo {author} {\bibfnamefont {I.~S.}\ \bibnamefont {Duff}}, \bibinfo {author} {\bibfnamefont {J.}~\bibnamefont {Koster}},\ and\ \bibinfo {author} {\bibfnamefont {J.-Y.}\ \bibnamefont {L'Excellent}},\ }\bibfield  {title} {\bibinfo {title} {A fully asynchronous multifrontal solver using distributed dynamic scheduling},\ }\href@noop {} {\bibfield  {journal} {\bibinfo  {journal} {SIAM Journal on Matrix Analysis and Applications}\ }\textbf {\bibinfo {volume} {23}},\ \bibinfo {pages} {15} (\bibinfo {year} {2001})}\BibitemShut {NoStop}%
\bibitem [{\citenamefont {Nangia}\ \emph {et~al.}(2019)\citenamefont {Nangia}, \citenamefont {Griffith}, \citenamefont {Patankar},\ and\ \citenamefont {Bhalla}}]{nangia2019robust}%
  \BibitemOpen
  \bibfield  {author} {\bibinfo {author} {\bibfnamefont {N.}~\bibnamefont {Nangia}}, \bibinfo {author} {\bibfnamefont {B.~E.}\ \bibnamefont {Griffith}}, \bibinfo {author} {\bibfnamefont {N.~A.}\ \bibnamefont {Patankar}},\ and\ \bibinfo {author} {\bibfnamefont {A.~P.~S.}\ \bibnamefont {Bhalla}},\ }\bibfield  {title} {\bibinfo {title} {{A robust incompressible Navier-Stokes solver for high density ratio multiphase flows}},\ }\href@noop {} {\bibfield  {journal} {\bibinfo  {journal} {Journal of Computational Physics}\ }\textbf {\bibinfo {volume} {390}},\ \bibinfo {pages} {548} (\bibinfo {year} {2019})}\BibitemShut {NoStop}%
\bibitem [{\citenamefont {Brown}\ \emph {et~al.}(2001)\citenamefont {Brown}, \citenamefont {Cortez},\ and\ \citenamefont {Minion}}]{brown2001accurate}%
  \BibitemOpen
  \bibfield  {author} {\bibinfo {author} {\bibfnamefont {D.~L.}\ \bibnamefont {Brown}}, \bibinfo {author} {\bibfnamefont {R.}~\bibnamefont {Cortez}},\ and\ \bibinfo {author} {\bibfnamefont {M.~L.}\ \bibnamefont {Minion}},\ }\bibfield  {title} {\bibinfo {title} {{Accurate projection methods for the incompressible Navier--Stokes equations}},\ }\href@noop {} {\bibfield  {journal} {\bibinfo  {journal} {Journal of computational physics}\ }\textbf {\bibinfo {volume} {168}},\ \bibinfo {pages} {464} (\bibinfo {year} {2001})}\BibitemShut {NoStop}%
\bibitem [{\citenamefont {Thirumalaisamy}\ \emph {et~al.}(2023)\citenamefont {Thirumalaisamy}, \citenamefont {Khedkar}, \citenamefont {Ghysels},\ and\ \citenamefont {Bhalla}}]{thirumalaisamy2023pre}%
  \BibitemOpen
  \bibfield  {author} {\bibinfo {author} {\bibfnamefont {R.}~\bibnamefont {Thirumalaisamy}}, \bibinfo {author} {\bibfnamefont {K.}~\bibnamefont {Khedkar}}, \bibinfo {author} {\bibfnamefont {P.}~\bibnamefont {Ghysels}},\ and\ \bibinfo {author} {\bibfnamefont {A.~P.~S.}\ \bibnamefont {Bhalla}},\ }\bibfield  {title} {\bibinfo {title} {An effective preconditioning strategy for volume penalized incompressible/low mach multiphase flow solvers},\ }\href {https://doi.org/https://doi.org/10.1016/j.jcp.2023.112325} {\bibfield  {journal} {\bibinfo  {journal} {Journal of Computational Physics}\ }\textbf {\bibinfo {volume} {490}},\ \bibinfo {pages} {112325} (\bibinfo {year} {2023})}\BibitemShut {NoStop}%
\bibitem [{IBA()}]{IBAMR-web-page}%
  \BibitemOpen
  \href@noop {} {\bibinfo {title} {{IBAMR}: {A}n adaptive and distributed-memory parallel implementation of the immersed boundary method}},\ \bibinfo {note} {\url{https://github.com/IBAMR/IBAMR}}\BibitemShut {NoStop}%
\bibitem [{\citenamefont {Hornung}\ and\ \citenamefont {Kohn}(2002)}]{HornungKohn02}%
  \BibitemOpen
  \bibfield  {author} {\bibinfo {author} {\bibfnamefont {R.~D.}\ \bibnamefont {Hornung}}\ and\ \bibinfo {author} {\bibfnamefont {S.~R.}\ \bibnamefont {Kohn}},\ }\bibfield  {title} {\bibinfo {title} {Managing application complexity in the {SAMRAI} object-oriented framework},\ }\href@noop {} {\bibfield  {journal} {\bibinfo  {journal} {Concurrency Comput Pract Ex}\ }\textbf {\bibinfo {volume} {14}},\ \bibinfo {pages} {347} (\bibinfo {year} {2002})}\BibitemShut {NoStop}%
\bibitem [{SAMRAI()}]{samrai-web-page}%
  \BibitemOpen
  SAMRAI,\ \href@noop {} {\bibinfo {title} {{SAMRAI}: {S}tructured {A}daptive {M}esh {R}efinement {A}pplication {I}nfrastructure}},\ \bibinfo {note} {\url{http://www.llnl.gov/CASC/SAMRAI}}\BibitemShut {NoStop}%
\bibitem [{\citenamefont {Balay}\ \emph {et~al.}(2015{\natexlab{a}})\citenamefont {Balay}, \citenamefont {Abhyankar}, \citenamefont {Adams}, \citenamefont {Brown}, \citenamefont {Brune}, \citenamefont {Buschelman}, \citenamefont {Dalcin}, \citenamefont {Eijkhout}, \citenamefont {Gropp}, \citenamefont {Kaushik}, \citenamefont {Knepley}, \citenamefont {McInnes}, \citenamefont {Rupp}, \citenamefont {Smith}, \citenamefont {Zampini},\ and\ \citenamefont {Zhang}}]{petsc-user-ref}%
  \BibitemOpen
  \bibfield  {author} {\bibinfo {author} {\bibfnamefont {S.}~\bibnamefont {Balay}}, \bibinfo {author} {\bibfnamefont {S.}~\bibnamefont {Abhyankar}}, \bibinfo {author} {\bibfnamefont {M.~F.}\ \bibnamefont {Adams}}, \bibinfo {author} {\bibfnamefont {J.}~\bibnamefont {Brown}}, \bibinfo {author} {\bibfnamefont {P.}~\bibnamefont {Brune}}, \bibinfo {author} {\bibfnamefont {K.}~\bibnamefont {Buschelman}}, \bibinfo {author} {\bibfnamefont {L.}~\bibnamefont {Dalcin}}, \bibinfo {author} {\bibfnamefont {V.}~\bibnamefont {Eijkhout}}, \bibinfo {author} {\bibfnamefont {W.~D.}\ \bibnamefont {Gropp}}, \bibinfo {author} {\bibfnamefont {D.}~\bibnamefont {Kaushik}}, \bibinfo {author} {\bibfnamefont {M.~G.}\ \bibnamefont {Knepley}}, \bibinfo {author} {\bibfnamefont {L.~C.}\ \bibnamefont {McInnes}}, \bibinfo {author} {\bibfnamefont {K.}~\bibnamefont {Rupp}}, \bibinfo {author} {\bibfnamefont {B.~F.}\ \bibnamefont {Smith}}, \bibinfo {author} {\bibfnamefont {S.}~\bibnamefont {Zampini}},\ and\ \bibinfo {author} {\bibfnamefont
  {H.}~\bibnamefont {Zhang}},\ }\href {http://www.mcs.anl.gov/petsc} {\emph {\bibinfo {title} {{PETS}c Users Manual}}},\ \bibinfo {type} {Tech. Rep.}\ \bibinfo {number} {ANL-95/11 - Revision 3.6}\ (\bibinfo  {institution} {Argonne National Laboratory},\ \bibinfo {year} {2015})\BibitemShut {NoStop}%
\bibitem [{\citenamefont {Balay}\ \emph {et~al.}(2015{\natexlab{b}})\citenamefont {Balay}, \citenamefont {Abhyankar}, \citenamefont {Adams}, \citenamefont {Brown}, \citenamefont {Brune}, \citenamefont {Buschelman}, \citenamefont {Dalcin}, \citenamefont {Eijkhout}, \citenamefont {Gropp}, \citenamefont {Kaushik}, \citenamefont {Knepley}, \citenamefont {McInnes}, \citenamefont {Rupp}, \citenamefont {Smith}, \citenamefont {Zampini},\ and\ \citenamefont {Zhang}}]{petsc-web-page}%
  \BibitemOpen
  \bibfield  {author} {\bibinfo {author} {\bibfnamefont {S.}~\bibnamefont {Balay}}, \bibinfo {author} {\bibfnamefont {S.}~\bibnamefont {Abhyankar}}, \bibinfo {author} {\bibfnamefont {M.~F.}\ \bibnamefont {Adams}}, \bibinfo {author} {\bibfnamefont {J.}~\bibnamefont {Brown}}, \bibinfo {author} {\bibfnamefont {P.}~\bibnamefont {Brune}}, \bibinfo {author} {\bibfnamefont {K.}~\bibnamefont {Buschelman}}, \bibinfo {author} {\bibfnamefont {L.}~\bibnamefont {Dalcin}}, \bibinfo {author} {\bibfnamefont {V.}~\bibnamefont {Eijkhout}}, \bibinfo {author} {\bibfnamefont {W.~D.}\ \bibnamefont {Gropp}}, \bibinfo {author} {\bibfnamefont {D.}~\bibnamefont {Kaushik}}, \bibinfo {author} {\bibfnamefont {M.~G.}\ \bibnamefont {Knepley}}, \bibinfo {author} {\bibfnamefont {L.~C.}\ \bibnamefont {McInnes}}, \bibinfo {author} {\bibfnamefont {K.}~\bibnamefont {Rupp}}, \bibinfo {author} {\bibfnamefont {B.~F.}\ \bibnamefont {Smith}}, \bibinfo {author} {\bibfnamefont {S.}~\bibnamefont {Zampini}},\ and\ \bibinfo {author} {\bibfnamefont
  {H.}~\bibnamefont {Zhang}},\ }\href {http://www.mcs.anl.gov/petsc} {\bibinfo {title} {{PETS}c {W}eb page}},\ \bibinfo {howpublished} {\url{http://www.mcs.anl.gov/petsc}} (\bibinfo {year} {2015}{\natexlab{b}})\BibitemShut {NoStop}%
\bibitem [{\citenamefont {Salari}\ and\ \citenamefont {Knupp}(2000)}]{salari2000code}%
  \BibitemOpen
  \bibfield  {author} {\bibinfo {author} {\bibfnamefont {K.}~\bibnamefont {Salari}}\ and\ \bibinfo {author} {\bibfnamefont {P.}~\bibnamefont {Knupp}},\ }\href@noop {} {\emph {\bibinfo {title} {Code verification by the method of manufactured solutions}}},\ \bibinfo {type} {Tech. Rep.}\ (\bibinfo  {institution} {Sandia National Lab.(SNL-NM), Albuquerque, NM (United States); Sandia~…},\ \bibinfo {year} {2000})\BibitemShut {NoStop}%
\bibitem [{\citenamefont {Barnkob}\ \emph {et~al.}(2018)\citenamefont {Barnkob}, \citenamefont {Nama}, \citenamefont {Ren}, \citenamefont {Huang}, \citenamefont {Costanzo},\ and\ \citenamefont {K{\"a}hler}}]{barnkob2018acoustically}%
  \BibitemOpen
  \bibfield  {author} {\bibinfo {author} {\bibfnamefont {R.}~\bibnamefont {Barnkob}}, \bibinfo {author} {\bibfnamefont {N.}~\bibnamefont {Nama}}, \bibinfo {author} {\bibfnamefont {L.}~\bibnamefont {Ren}}, \bibinfo {author} {\bibfnamefont {T.~J.}\ \bibnamefont {Huang}}, \bibinfo {author} {\bibfnamefont {F.}~\bibnamefont {Costanzo}},\ and\ \bibinfo {author} {\bibfnamefont {C.~J.}\ \bibnamefont {K{\"a}hler}},\ }\bibfield  {title} {\bibinfo {title} {Acoustically driven fluid and particle motion in confined and leaky systems},\ }\href@noop {} {\bibfield  {journal} {\bibinfo  {journal} {Physical Review Applied}\ }\textbf {\bibinfo {volume} {9}},\ \bibinfo {pages} {014027} (\bibinfo {year} {2018})}\BibitemShut {NoStop}%
\bibitem [{\citenamefont {Lighthill}(1978)}]{lighthill1978acoustic}%
  \BibitemOpen
  \bibfield  {author} {\bibinfo {author} {\bibfnamefont {J.}~\bibnamefont {Lighthill}},\ }\bibfield  {title} {\bibinfo {title} {Acoustic streaming},\ }\href@noop {} {\bibfield  {journal} {\bibinfo  {journal} {Journal of sound and vibration}\ }\textbf {\bibinfo {volume} {61}},\ \bibinfo {pages} {391} (\bibinfo {year} {1978})}\BibitemShut {NoStop}%
\bibitem [{\citenamefont {Chini}\ \emph {et~al.}(2014)\citenamefont {Chini}, \citenamefont {Malecha},\ and\ \citenamefont {Dreeben}}]{chini2014large}%
  \BibitemOpen
  \bibfield  {author} {\bibinfo {author} {\bibfnamefont {G.}~\bibnamefont {Chini}}, \bibinfo {author} {\bibfnamefont {Z.}~\bibnamefont {Malecha}},\ and\ \bibinfo {author} {\bibfnamefont {T.}~\bibnamefont {Dreeben}},\ }\bibfield  {title} {\bibinfo {title} {Large-amplitude acoustic streaming},\ }\href@noop {} {\bibfield  {journal} {\bibinfo  {journal} {Journal of fluid mechanics}\ }\textbf {\bibinfo {volume} {744}},\ \bibinfo {pages} {329} (\bibinfo {year} {2014})}\BibitemShut {NoStop}%
\bibitem [{\citenamefont {Dillinger}\ \emph {et~al.}(2024)\citenamefont {Dillinger}, \citenamefont {Knipper}, \citenamefont {Nama},\ and\ \citenamefont {Ahmed}}]{dillinger2024steerable}%
  \BibitemOpen
  \bibfield  {author} {\bibinfo {author} {\bibfnamefont {C.}~\bibnamefont {Dillinger}}, \bibinfo {author} {\bibfnamefont {J.}~\bibnamefont {Knipper}}, \bibinfo {author} {\bibfnamefont {N.}~\bibnamefont {Nama}},\ and\ \bibinfo {author} {\bibfnamefont {D.}~\bibnamefont {Ahmed}},\ }\bibfield  {title} {\bibinfo {title} {Steerable acoustically powered starfish-inspired microrobot},\ }\href@noop {} {\bibfield  {journal} {\bibinfo  {journal} {Nanoscale}\ }\textbf {\bibinfo {volume} {16}},\ \bibinfo {pages} {1125} (\bibinfo {year} {2024})}\BibitemShut {NoStop}%
\bibitem [{\citenamefont {Ren}\ \emph {et~al.}(2019)\citenamefont {Ren}, \citenamefont {Nama}, \citenamefont {McNeill}, \citenamefont {Soto}, \citenamefont {Yan}, \citenamefont {Liu}, \citenamefont {Wang}, \citenamefont {Wang},\ and\ \citenamefont {Mallouk}}]{ren20193d}%
  \BibitemOpen
  \bibfield  {author} {\bibinfo {author} {\bibfnamefont {L.}~\bibnamefont {Ren}}, \bibinfo {author} {\bibfnamefont {N.}~\bibnamefont {Nama}}, \bibinfo {author} {\bibfnamefont {J.~M.}\ \bibnamefont {McNeill}}, \bibinfo {author} {\bibfnamefont {F.}~\bibnamefont {Soto}}, \bibinfo {author} {\bibfnamefont {Z.}~\bibnamefont {Yan}}, \bibinfo {author} {\bibfnamefont {W.}~\bibnamefont {Liu}}, \bibinfo {author} {\bibfnamefont {W.}~\bibnamefont {Wang}}, \bibinfo {author} {\bibfnamefont {J.}~\bibnamefont {Wang}},\ and\ \bibinfo {author} {\bibfnamefont {T.~E.}\ \bibnamefont {Mallouk}},\ }\bibfield  {title} {\bibinfo {title} {3d steerable, acoustically powered microswimmers for single-particle manipulation},\ }\href@noop {} {\bibfield  {journal} {\bibinfo  {journal} {Science advances}\ }\textbf {\bibinfo {volume} {5}},\ \bibinfo {pages} {eaax3084} (\bibinfo {year} {2019})}\BibitemShut {NoStop}%
\bibitem [{\citenamefont {Ahmed}\ \emph {et~al.}(2016)\citenamefont {Ahmed}, \citenamefont {Ozcelik}, \citenamefont {Bojanala}, \citenamefont {Nama}, \citenamefont {Upadhyay}, \citenamefont {Chen}, \citenamefont {Hanna-Rose},\ and\ \citenamefont {Huang}}]{ahmed2016rotational}%
  \BibitemOpen
  \bibfield  {author} {\bibinfo {author} {\bibfnamefont {D.}~\bibnamefont {Ahmed}}, \bibinfo {author} {\bibfnamefont {A.}~\bibnamefont {Ozcelik}}, \bibinfo {author} {\bibfnamefont {N.}~\bibnamefont {Bojanala}}, \bibinfo {author} {\bibfnamefont {N.}~\bibnamefont {Nama}}, \bibinfo {author} {\bibfnamefont {A.}~\bibnamefont {Upadhyay}}, \bibinfo {author} {\bibfnamefont {Y.}~\bibnamefont {Chen}}, \bibinfo {author} {\bibfnamefont {W.}~\bibnamefont {Hanna-Rose}},\ and\ \bibinfo {author} {\bibfnamefont {T.~J.}\ \bibnamefont {Huang}},\ }\bibfield  {title} {\bibinfo {title} {Rotational manipulation of single cells and organisms using acoustic waves},\ }\href@noop {} {\bibfield  {journal} {\bibinfo  {journal} {Nature communications}\ }\textbf {\bibinfo {volume} {7}},\ \bibinfo {pages} {11085} (\bibinfo {year} {2016})}\BibitemShut {NoStop}%
\bibitem [{\citenamefont {Guo}\ \emph {et~al.}(2015)\citenamefont {Guo}, \citenamefont {Li}, \citenamefont {French}, \citenamefont {Mao}, \citenamefont {Zhao}, \citenamefont {Li}, \citenamefont {Nama}, \citenamefont {Fick}, \citenamefont {Benkovic},\ and\ \citenamefont {Huang}}]{guo2015controlling}%
  \BibitemOpen
  \bibfield  {author} {\bibinfo {author} {\bibfnamefont {F.}~\bibnamefont {Guo}}, \bibinfo {author} {\bibfnamefont {P.}~\bibnamefont {Li}}, \bibinfo {author} {\bibfnamefont {J.~B.}\ \bibnamefont {French}}, \bibinfo {author} {\bibfnamefont {Z.}~\bibnamefont {Mao}}, \bibinfo {author} {\bibfnamefont {H.}~\bibnamefont {Zhao}}, \bibinfo {author} {\bibfnamefont {S.}~\bibnamefont {Li}}, \bibinfo {author} {\bibfnamefont {N.}~\bibnamefont {Nama}}, \bibinfo {author} {\bibfnamefont {J.~R.}\ \bibnamefont {Fick}}, \bibinfo {author} {\bibfnamefont {S.~J.}\ \bibnamefont {Benkovic}},\ and\ \bibinfo {author} {\bibfnamefont {T.~J.}\ \bibnamefont {Huang}},\ }\bibfield  {title} {\bibinfo {title} {Controlling cell--cell interactions using surface acoustic waves},\ }\href@noop {} {\bibfield  {journal} {\bibinfo  {journal} {Proceedings of the National Academy of Sciences}\ }\textbf {\bibinfo {volume} {112}},\ \bibinfo {pages} {43} (\bibinfo {year} {2015})}\BibitemShut {NoStop}%
\bibitem [{\citenamefont {Li}\ \emph {et~al.}(2015)\citenamefont {Li}, \citenamefont {Ding}, \citenamefont {Mao}, \citenamefont {Chen}, \citenamefont {Nama}, \citenamefont {Guo}, \citenamefont {Li}, \citenamefont {Wang}, \citenamefont {Cameron},\ and\ \citenamefont {Huang}}]{li2015standing}%
  \BibitemOpen
  \bibfield  {author} {\bibinfo {author} {\bibfnamefont {S.}~\bibnamefont {Li}}, \bibinfo {author} {\bibfnamefont {X.}~\bibnamefont {Ding}}, \bibinfo {author} {\bibfnamefont {Z.}~\bibnamefont {Mao}}, \bibinfo {author} {\bibfnamefont {Y.}~\bibnamefont {Chen}}, \bibinfo {author} {\bibfnamefont {N.}~\bibnamefont {Nama}}, \bibinfo {author} {\bibfnamefont {F.}~\bibnamefont {Guo}}, \bibinfo {author} {\bibfnamefont {P.}~\bibnamefont {Li}}, \bibinfo {author} {\bibfnamefont {L.}~\bibnamefont {Wang}}, \bibinfo {author} {\bibfnamefont {C.~E.}\ \bibnamefont {Cameron}},\ and\ \bibinfo {author} {\bibfnamefont {T.~J.}\ \bibnamefont {Huang}},\ }\bibfield  {title} {\bibinfo {title} {Standing surface acoustic wave (ssaw)-based cell washing},\ }\href@noop {} {\bibfield  {journal} {\bibinfo  {journal} {Lab on a Chip}\ }\textbf {\bibinfo {volume} {15}},\ \bibinfo {pages} {331} (\bibinfo {year} {2015})}\BibitemShut {NoStop}%
\bibitem [{\citenamefont {Ozcelik}\ \emph {et~al.}(2016)\citenamefont {Ozcelik}, \citenamefont {Nama}, \citenamefont {Huang}, \citenamefont {Kaynak}, \citenamefont {McReynolds}, \citenamefont {Hanna-Rose},\ and\ \citenamefont {Huang}}]{ozcelik2016acoustofluidic}%
  \BibitemOpen
  \bibfield  {author} {\bibinfo {author} {\bibfnamefont {A.}~\bibnamefont {Ozcelik}}, \bibinfo {author} {\bibfnamefont {N.}~\bibnamefont {Nama}}, \bibinfo {author} {\bibfnamefont {P.-H.}\ \bibnamefont {Huang}}, \bibinfo {author} {\bibfnamefont {M.}~\bibnamefont {Kaynak}}, \bibinfo {author} {\bibfnamefont {M.~R.}\ \bibnamefont {McReynolds}}, \bibinfo {author} {\bibfnamefont {W.}~\bibnamefont {Hanna-Rose}},\ and\ \bibinfo {author} {\bibfnamefont {T.~J.}\ \bibnamefont {Huang}},\ }\bibfield  {title} {\bibinfo {title} {Acoustofluidic rotational manipulation of cells and organisms using oscillating solid structures},\ }\href@noop {} {\bibfield  {journal} {\bibinfo  {journal} {Small (Weinheim an der Bergstrasse, Germany)}\ }\textbf {\bibinfo {volume} {12}},\ \bibinfo {pages} {5120} (\bibinfo {year} {2016})}\BibitemShut {NoStop}%
\bibitem [{\citenamefont {Thirumalaisamy}\ and\ \citenamefont {Bhalla}(2023)}]{thirumalaisamy2023lowmach}%
  \BibitemOpen
  \bibfield  {author} {\bibinfo {author} {\bibfnamefont {R.}~\bibnamefont {Thirumalaisamy}}\ and\ \bibinfo {author} {\bibfnamefont {A.~P.~S.}\ \bibnamefont {Bhalla}},\ }\bibfield  {title} {\bibinfo {title} {A low mach enthalpy method to model non-isothermal gas–liquid–solid flows with melting and solidification},\ }\href@noop {} {\bibfield  {journal} {\bibinfo  {journal} {International Journal of Multiphase Flow}\ }\textbf {\bibinfo {volume} {169}},\ \bibinfo {pages} {104605} (\bibinfo {year} {2023})}\BibitemShut {NoStop}%
\bibitem [{\citenamefont {Thirumalaisamy}\ and\ \citenamefont {Bhalla}(2025)}]{thirumalaisamy2025consistent}%
  \BibitemOpen
  \bibfield  {author} {\bibinfo {author} {\bibfnamefont {R.}~\bibnamefont {Thirumalaisamy}}\ and\ \bibinfo {author} {\bibfnamefont {A.~P.~S.}\ \bibnamefont {Bhalla}},\ }\bibfield  {title} {\bibinfo {title} {A consistent, volume preserving, and adaptive mesh refinement-based framework for modeling non-isothermal gas--liquid--solid flows with phase change},\ }\href@noop {} {\bibfield  {journal} {\bibinfo  {journal} {International Journal of Multiphase Flow}\ }\textbf {\bibinfo {volume} {183}},\ \bibinfo {pages} {105060} (\bibinfo {year} {2025})}\BibitemShut {NoStop}%
\end{thebibliography}%

\end{document}